\documentclass[acmsmall]{acmart}

\AtBeginDocument{%
  \providecommand\BibTeX{{%
    \normalfont B\kern-0.5em{\scshape i\kern-0.25em b}\kern-0.8em\TeX}}}

\usepackage{booktabs} % For formal tables
\usepackage{enumitem}
\usepackage{mathtools}
\usepackage{dsfont}
\usepackage{amsmath}
\usepackage{amssymb}
\usepackage[linesnumbered,boxed,ruled]{algorithm2e}
\usepackage{graphicx, subfigure}
\usepackage[perpage,symbol]{footmisc}
\setcopyright{acmcopyright}
\setcopyright{acmcopyright}
\acmJournal{TIST}
\acmYear{2020} \acmVolume{1} \acmNumber{1} \acmArticle{1} \acmMonth{1} \acmPrice{15.00}\acmDOI{10.1145/3394138}

\begin{document}
\title{Practical Privacy Preserving POI Recommendation}
%\titlenote{Produces the permission block, and
%copyright information}
%\subtitle{Extended Abstract}
%\subtitlenote{The full version of the author's guide is available as
%\texttt{acmart.pdf} document}

\author{Chaochao Chen}
\affiliation{%
  \institution{Ant Financial Services Group}
  \country{China}}
\email{chaochao.ccc@antfin.com}

\author{Jun Zhou}
\affiliation{%
  \institution{Ant Financial Services Group}
  \country{China}}
\email{jun.zhoujun@antfin.com}

\author{Bingzhe Wu}
\affiliation{%
  \institution{Peking University}
  \country{China}}
\email{wubingzhe@pku.edu.cn}

\author{Wenjing Fang}
\affiliation{%
  \institution{Ant Financial Services Group}
  \country{China}}
\email{bean.fwj@antfin.com}

\author{Li Wang}
\affiliation{%
  \institution{Ant Financial Services Group}
  \country{China}}
\email{raymond.wangl@antfin.com}

\author{Yuan Qi}
\affiliation{%
  \institution{Ant Financial Services Group}
  \country{China}}
\email{yuan.qi@antfin.com}

\author{Xiaolin Zheng*}\footnote{Corresponding author. This work was supported in part by the National Key R\&D Program of China (No. 2018YFB1403001). }
\affiliation{%
  \institution{Zhejiang University}
  \country{China}}
\email{xlzheng@zju.edu.cn}

%\author{Chaochao Chen, Ziqi Liu, Li Wang, Wenjing Fang, Liang Li, Jun Zhou, Yuan Qi}
%\affiliation{%
%  \institution{AI Department, Ant Financial Services Group}
%  \streetaddress{P.O. Box 1212}
%  \city{Hangzhou}
%  \state{China}
%  \postcode{310000}
%}
%\email{{chaochao.ccc, ziqiliu, raymond.wangl, bean.fwj, liangli.ll, jun.zhoujun, yuan.qi}@antfin.com}

\renewcommand{\shortauthors}{Chaochao Chen et al.}
\newcommand{\tabincell}[2]{\begin{tabular}{@{}#1@{}}#2\end{tabular}}

\begin{abstract}
%Recommender system has been drawing much attention in recent decades, and achieving great success in many real-world applications such as e-commerce and Point-of-Interest (POI) recommendation. 
Point-of-Interest (POI) recommendation has been extensively studied and successfully applied in industry recently. 
However, most existing approaches build centralized models on the basis of collecting users' data. 
Both private data and models are held by the recommender, which causes serious privacy concerns. 
In this paper, we propose a novel \texttt{Pri}vacy preserving POI \texttt{Rec}ommendation (\texttt{PriRec}) framework. 
First, to protect data privacy, users' private data (features and actions) are kept on their own side, e.g., Cellphone or Pad. 
Meanwhile, the public data need to be accessed by all the users are kept by the recommender to reduce the storage costs of users' devices. 
Those public data include: (1) static data only related to the status of POI, such as POI categories, and (2) dynamic data depend on user-POI actions such as visited counts. 
The dynamic data could be sensitive, and we develop local differential privacy techniques to release such data to public with privacy guarantees.
Second, \texttt{PriRec} follows the representations of Factorization Machine (FM) that consists of linear model and the feature interaction model. 
To protect the model privacy, the linear models are saved on users' side, and we propose a secure decentralized gradient descent protocol for users to learn it collaboratively. 
The feature interaction model is kept by the recommender since there is no privacy risk, and we adopt secure aggregation strategy in federated learning paradigm to learn it. 
To this end, \texttt{PriRec} keeps users' private raw data and models in users' own hands, and protects user privacy to a large extent. 
We apply \texttt{PriRec} in real-world datasets, and comprehensive experiments demonstrate that, compared with FM, \texttt{PriRec} achieves comparable or even better recommendation accuracy. 
\end{abstract}

\begin{CCSXML}
<ccs2012>
<concept>
<concept_id>10002951.10003317.10003338</concept_id>
<concept_desc>Information systems~Retrieval models and ranking</concept_desc>
<concept_significance>500</concept_significance>
</concept>
<concept>
<concept_id>10002978.10003029.10011150</concept_id>
<concept_desc>Security and privacy~Privacy protections</concept_desc>
<concept_significance>500</concept_significance>
</concept>
</ccs2012>
\end{CCSXML}

\ccsdesc[500]{Information systems~Retrieval models and ranking}
\ccsdesc[500]{Security and privacy~Privacy protections}

\keywords{privacy preserving, decentralization, local differential privacy, secret sharing, POI recommendation}

\maketitle

\section{Introduction}	

Recommender system has been drawing much attention in recent decades, and achieving great successes in many real-world applications such as vedio \cite{covington2016deep}, e-commerce \cite{Wang2018BCE}, and Point-of-Interest (POI) (e.g., restaurant and hotel) recommendation \cite{yang2017bridging}, to solve the information overload problem. 
Take POI recommendation as an example, most promising models are centralizedly built on the basis of collecting users' private data, which causes serious privacy concerns \cite{lam2006you,mcsherry2009differentially,riboni2012private}.

\begin{figure}[t]
\centering
\includegraphics[width=8cm]{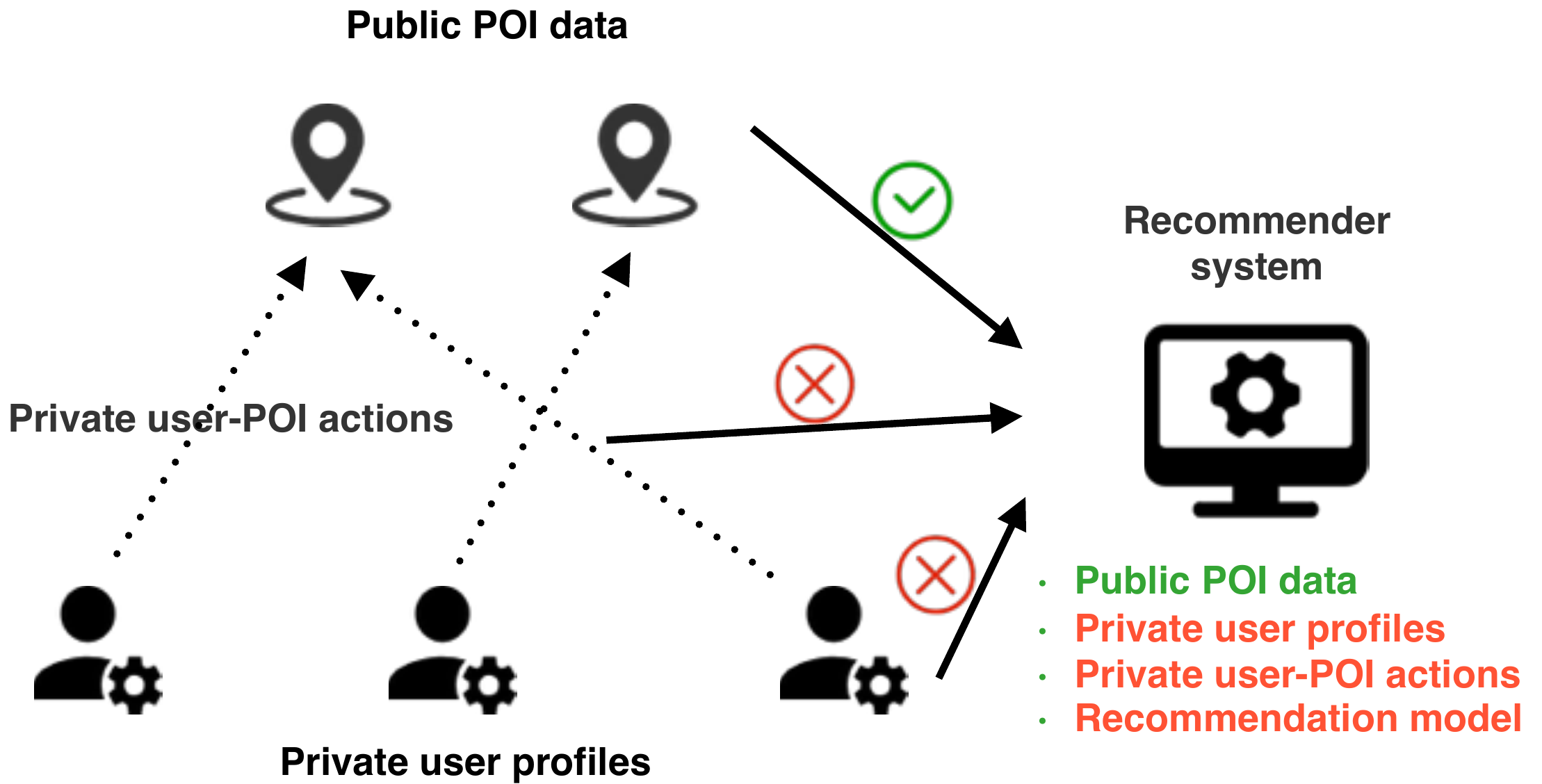}
%\vskip -0.05in
\caption{Traditional POI recommendation framework. Users' private data and models are centralizedly kept by the recommender, which raise serious privacy concerns.}
\label{toyexample}
%\vskip -0.1in
\end{figure}

\textbf{A motivating example. }
Figure \ref{toyexample} shows the framework of most existing POI recommendation approaches, where the data include user profiles (e.g., age and gender), POI descriptions (e.g., category and visited count), and user-POI actions (e.g., click and check-in). 
Among them, both user profiles and user-POI actions are private, whereas POI descriptions are public to all the users. 
The model refers to the built recommendation model, e.g., the latent factors of Matrix Factorization (MF) model \cite{koren2008factorization}, that predicts users' preferences on POIs. 
%Since both data and model are kept by the recommender, it raises serious privacy concerns. 
First, besides the public POI data, users' private data, including user profiles and user-POI actions, are collected, and these data explicitly show users' private information and may be abused by the recommender. 
Second, the models of most existing POI recommendation approaches implicitly indicate users' private information, e.g., the latent factors of MF can directly infer users' ratings on items. 
Therefore, both data and models of most existing recommender systems could be in high privacy risks \cite{polat2003privacy,ricci2015recommender}. 
%How to build a privacy preserving POI recommender system has been a challenging task \cite{ricci2015recommender}. 

There have been some studies focus on protecting user privacy while building recommender systems, including the applications to POI recommendation \cite{riboni2012private,chen2018privacy}. 
They mainly belong to two types. 
The first type protects the raw data by adding noises to them \cite{polat2005privacy,berkovsky2007enhancing,mcsherry2009differentially,riboni2012private,hua2015differentially,meng2018personalized}. 
%The adopted techniques are mainly \textit{randomized perturbation} and \textit{differential privacy} \cite{dwork2008differential}, and popular methods of this type include \cite{polat2005privacy,berkovsky2007enhancing,mcsherry2009differentially,riboni2012private,hua2015differentially,meng2018personalized}. 
These methods are efficient and easy to implement, however, the recommendation performance decreases when adding too much noises. 
The second type is based on cryptography techniques \cite{canny2002collaborative,polat2005privacy,aimeur2008lambic,erkin2010privacy,nikolaenko2013privacy}. 
These approaches usually can achieve comparable performance with the traditional recommender systems, however, their efficiencies are too low to be applied in practice. 
%Both types of approaches need to collect users' private data, either perturbed data or encrypted data, which is less user-friendly compared with do not collect users' private data at all. 
Therefore, how to build a privacy preserving recommender system, which can not only protect user data and model privacy, but also has comparable (or even better) recommendation accuracy and high efficiency, remains a challenge.

To solve above challenges, in this paper, we take POI recommendation as a Click-Through Rate (CTR) prediction problem, and propose a novel \texttt{Pri}vacy preserving POI \texttt{Rec}ommendation (\texttt{PriRec}) framework, which has the following advantages. 

\textbf{\texttt{PriRec} protects data and model privacy. }
First, \textit{to protect data privacy}, \texttt{PriRec} keeps users' private data (features and actions) on their own side, e.g., Cellphone or Pad. 
To alleviate the storage costs on users' devices, all the public POIs' data are still held by the recommender. 
These public data can be divided into two types: 
(1) the static POI data that describe the status of a POI such as POI categories, and (2) the dynamic POI data that indicate the popularity of a POI, e.g., visited count. 
Since the user-POI actions are kept on users' devices, to obtain the statistics of these action data, we propose to use local differential privacy technique \cite{ding2017collecting} to collect perturbed user-POI interaction data, and further generate POI dynamic features by the recommender. 
Here, different from the existing models \cite{hua2015differentially}, which directly use the perturbed data to build models, we only use the perturbed user-POI interaction data for generating statistical features. 
Second, \textit{to protect model privacy}, motivated by Factorization Machine (FM) \cite{rendle2012factorization}, we design the model of \texttt{PriRec} as two parts: 
(1) the linear models that are decentralized on each user's side since they directly indicate user preferences, 
and (2) the feature interaction model that is kept by the recommender has no privacy risk since it can only infer the interaction weights between features.
To this end, both users' private raw data and models are kept by their own hands, and \texttt{PriRec} is able to protect user privacy to a large extent. 

\textbf{\texttt{PriRec} has linear time complexity and promising recommendation accuracy. }
The learning process in \texttt{PriRec} includes two parts, the learning of linear models on each user's side and the learning of feature interaction model kept by the recommender. 
First, inspired by decentralized gradient descent \cite{nedic2009distributed,yuan2016convergence}, we propose a \textit{secure decentralized gradient descent protocol} for users to learn their linear models collaboratively. 
Second, motivated by parameter server distributed learning paradigm \cite{li2014scaling} and federated learning \cite{konevcny2016federated,bonawitz2017practical}, we adopt \textit{secure aggregation strategy in federated learning paradigm} to learn the feature interaction model. 
Both strategies are efficient and make the learning of \texttt{PriRec} scales linearly with data size in terms of both computation and communication complexities. 
Moreover, \texttt{PriRec} belongs to decentralized model and it learns the linear models for different users based on location networks. 
To this end, PriRec can capture users' individual interests in different locations, and achieve promising recommendation accuracy. 

We apply \texttt{PriRec} in real-world datasets, and comprehensive experiments demonstrate that, compared with the traditional ranking model, \texttt{PriRec} achieves comparable or even better recommendation performance, and meanwhile keeps user privacy. 

Our main contributions are summarized as follows:
\begin{itemize}% [leftmargin=*] \setlength{\itemsep}{-\itemsep}
\item We propose a novel \texttt{Pri}vacy preserving POI \texttt{Rec}ommendation (\texttt{PriRec}) framework for POI recommendation, where we propose a secure decentralized gradient descent protocol for learning decentralized linear models and adopt secure aggregation strategy in federated learning paradigm to learn the feature interaction model. 
\texttt{PriRec} keeps users' private raw data and models on users' own side, and therefore protects user privacy to a large extent. 
\item We propose to adopt local differential privacy techniques to generate dynamic POI popularity features from users' local user-POI actions. This can not only protect private user-POI actions, but also significantly improve recommendation performance, as we will show in experiments. 
% \item We propose a secure decentralized gradient descent protocol for learning decentralized linear models, and adopt secure aggregation strategy in federated learning paradigm to learn the feature interaction model. 
% The complexity analysis and empirical study both show that our proposed model learning algorithm scales linearly with data size in terms of both computation and communication complexities. 
\item We conduct experiments on real-world datasets, and the results demonstrate the effectiveness and efficiency of \texttt{PriRec}.
\end{itemize}

\section{Related Work}\label{background}
In this section, we review related knowledge, including the traditional recommender system, privacy preserving recommender system, local differential privacy, and secret sharing.

\subsection{Traditional Recommender System}
We first review literatures of traditional recommender system, i.e., non-privacy preserving approaches, including the applications in POI recommendations. 
The most famous traditional recommender system is Collaborative Filtering (CF) \cite{sarwar2001item,su2009survey,ye2011exploiting}, which is based on the assumption that users who behave similarly on some items will also behave similarly on other items. 
Among CF, factorization based models achieve promising performance \cite{koren2008factorization,li2015rank,chen2018semi}, which aim to learn user and item latent factors based on known user-item action histories such as ratings and clicks. 
%The output of factorization based models ensure that users who have similar behaviors have similar latent factors. 
Popular factorization based CF models include Matrix Factorization (MF) and its variants \cite{mnih2007probabilistic,koren2008factorization,cheng2012fused,yang2013sentiment,lian2014geomf}, regression-based latent factor models \cite{agarwal2009regression}, Bayesian personalized ranking \cite{rendle2009bpr}, deep MF \cite{xue2017deep}, neural MF \cite{he2017neural}, and Hash based MF \cite{Chen2018DCH}. 

Besides the above CF models, in practice, ads, merchandise, and POI recommendations are also taken as a Click-Through Rate (CTR) prediction problem \cite{ling2017model}. 
Logistic Regression (LR) is popularly used in most Internet companies, e.g., Microsoft \cite{richardson2007predicting}, Google \cite{mcmahan2013ad}, and Ant Financial \cite{chen2017large}, due to its simplicity, scalability, and online learning capability. 
Deep neural network (DNN) has also been widely used due to its powerful representation ability \cite{zhang2016deep,zhu2018deep}. 
Later on, Wide \& deep \cite{cheng2016wide} combines the advantages of both LR and DNN for better performance. 
Besides, Factorization Machine (FM) \cite{rendle2012factorization} and its variations, e.g., DeepFM \cite{guo2017deepfm} and Field-aware FM \cite{juan2016field}, are also extensively used since they can capture the high-order interactions between features. 

Although the traditional recommender systems achieve promising performance, they build centralized recommendation models on the basis of collecting users' data. 
Both private data (features and actions) and models are hold by the recommender, which cause serious privacy concerns \cite{lam2006you,vallet2014matrix,ricci2015recommender}. 
In this paper, we take POI recommendation as a CTR prediction problem, and propose a novel \texttt{Pri}vacy preserving POI \texttt{Rec}ommendation (\texttt{PriRec}) framework for it. 
\texttt{PriRec} keeps users' private data and models on users' own side, e.g., Cellphone or Pad, thus solves the privacy issue. 
% All the POIs' data are kept by the recommender for saving storage cost, since there is no user privacy leakage. 
% We also propose a private Factorization Machine (FM) model, which includes two parts: 
% (1) the linear models are saved on each user's side for privacy purpose, and we propose a model propagation strategy for users to learn their linear models collaboratively, 
% and (2) the feature interaction model is kept by the recommender since there is no privacy risk, and we propose a parameter-server like communication strategy to learn it. 

\subsection{Privacy Preserving Recommender System}
%The traditional recommender systems raise serious privacy concerns \cite{lam2006you,ricci2015recommender}. 
To date, different approaches have been proposed to solve the privacy issues of the traditional recommender systems. 
The first type is based on randomized perturbation or differential privacy techniques \cite{dwork2008differential}. 
That is, they protect users' original data by adding noise to them. 
Popular methods of this type include \cite{polat2005privacy,berkovsky2007enhancing,mcsherry2009differentially,riboni2012private,hua2015differentially,meng2018personalized}. 
These methods are efficient and easy to implement, however, there is a trade-off bewteen privacy and recommendation accuracy, i.e., the recommendation performance decreases when the privacy degree increases. 
The second type is based on cryptography techniques such as homomorphic encryption \cite{gentry2009fully} and secure Multi-Party Computation (MPC) \cite{yao1986generate}, and typical methods include \cite{canny2002collaborative,polat2005privacy,aimeur2008lambic,erkin2010privacy,nikolaenko2013privacy}. 
These approaches usually can achieve comparable performance with the traditional recommender systems, however, the low efficiency of the cryptography techniques limits its application in practice. 

Besides the above privacy-preserving recommendation models, there are also existing approaches focus on combining the private data of multi-parties, e.g., different hospitals and banks, meanwhile training machine learning models such as LR \cite{chaudhuri2009privacy,mohassel2017secureml}, which is the so-called \textit{collaborative learning} or \textit{shared machine learning} in literature \cite{chen2020secure}. 
They do this by using differential privacy or MPC. 
The fundamental difference between these works and ours is that, they assume users' data have been collected by several parties who want to protect their collected data from other parties, while our approach assumes users' private raw data are kept on their own devices. 

The most similar work to ours is Federated Learning (FL) \cite{konevcny2016federated,bonawitz2017practical}. %and Decentralized Matrix Factorization (DMF) \cite{chen2018privacy}. 
However, \texttt{PriRec} is different from FL in two aspects: 
(1) FL assumes that data are decentralized on each user's device and the model is kept by the server (recommender), while in \texttt{PriRec}, both users' private data and models are decentralized on each user's device, and therefore \texttt{PriRec} has better user privacy guarantees; 
(2) FL only uses secure gradient aggregation strategy to learn neural network model while \texttt{PriRec} uses both secure decentralized gradient descent protocol and secure gradient aggregation strategy to learn FM model. 
% Second, \texttt{PriRec} differs from DMF mainly in two aspects: 
% (1) DMF can only handle user-item rating information, in contrast, \texttt{PriRec} is able to take both user-item rating and feature information into account. This is similar as the difference between MF and FM, and it has been proven that FM significantly outperforms MF \cite{rendle2012factorization}. 
% (2) As for model security, it needs to send out model gradient to learn the DMF model. However, existing research has proven that it is possible to obtain the private training data from the publicly shared gradients \cite{zhu2019deep}. 
% In comparison, in \texttt{PriRec}, we propose a secure decentralized gradient descent protocol based on secret sharing technique \cite{shamir1979share}, with which both model and gradient are securely protected. 

\subsection{Local Differential Privacy}
Differential Privacy (DP) has been proposed in the global privacy context to ensure that an adversary should not be able to reliably infer whether or not a particular individual is participating in the database query, while Local Differential Privacy (LDP) was proposed in the local privacy context, as in when individuals disclose their personal information \cite{kairouz2014extremal,cormode2018privacy}. 
LDP has the ability of estimating statistical values of data, e.g., mean and histogram, without disclose users' raw data, and has been adopted by many companies, including Google \cite{erlingsson2014rappor}, Apple \cite{thakurta2017learning}, and Microsoft \cite{ding2017collecting}. 
Recently, LDP has also been applied in recommender system to protect private user-item ratings \cite{shen2016epicrec,shin2018privacy}. 
However, directly using LDP to build models will decrease recommendation performance. 

In this paper, we propose to adopt LDP to generate dynamic POI features (e.g., the visited count of a POI) instead of directly building models, which can protect user-POI actions and capture the popularity of POIs. 
We will show in experiments that the generated POI features can significantly improve recommendation performance. 

\subsection{Secret Sharing}
Secret sharing was first proposed in \cite{shamir1979share}. 
The basic idea of secret sharing is to distribute a secret amongst a group of participants (parties), each of whom has a share of the secret. 
The secret can be reconstructed only when a sufficient number of shares are combined together, and individual shares are of no use on their own.
We focus on $n$-out-of-$n$ Secret Sharing in this paper, i.e., all shares are needed to reconstruct a secret. 
To share an $\ell$-bit value $a$ for party $i \in \mathcal{P} = \{1,...,P\}$, party $i$ generates \{$a_j \in \mathds{Z}_{2^\ell}, j \in \mathcal{P}$ and $j \neq i$\} uniformly at random, sends $a_j$ to to party $j$, and keeps $a_i = a- \sum_j a_j$ mod $2^\ell$. 
We use $\langle a \rangle _i = a_i$ to denote the share of party $i$. To reconstruct a shared value $\langle a \rangle$, each party $i$ sends $\langle a \rangle _i$ to one who computes $\sum_i a_i$ mod $2^\ell, i \in \mathcal{P}$. 

The above protocols can not work directly with decimal numbers, since it is not possible to sample uniformly in $\mathds{R}$ \cite{cock2015fast}. We approximate decimal arithmetics following the existing work \cite{mohassel2017secureml}. Suppose $a$ and $b$ are two decimal numbers with at most $l_F$ bits in the fractional part, to do fixed-point multiplication, we first transform them to integers by letting $a'=2^{l_F}a$ and $b'=2^{l_F}b$, and then calculate $z=a'b'$. Finally, we truncate the last $l_F$ bits of $z$ so that it has at most $l_F$ bits representing the fractional part. It has been proven that this truncation technique also works when $z$ is secret shared \cite{mohassel2017secureml}. 

Secret sharing has been popolarly used in kinds of machine learning algorithms, including linear regression \cite{cock2015fast}, neural network \cite{mohassel2017secureml}, and recommender system \cite{chen2020secure}. In this paper, we apply secret sharing into decentralized gradient descent, and propose a secure decentralized gradient descent protocol for users to learn the linear model of \texttt{PriRec} collaboratively, without compromising users' private data and model. 

\section{The Proposed Privacy Preserving POI Recommendation Framework}
In this section, we first describe motivations, notations, and problem definitions. 
Next, we present the \texttt{Pri}vacy preserving POI \texttt{Rec}ommendation (\texttt{PriRec}) framework, followed by its main components in details. 
We then summarize the training and prediction algorithms of \texttt{PriRec}, and finally analyze their complexities.

\subsection{Preliminary}
We first describe the motivation of our proposed \texttt{PriRec} framework, and then present the notations and problem definition, and finally describe model optimization. 

\begin{figure*}[th]
\centering
\includegraphics[width=13cm]{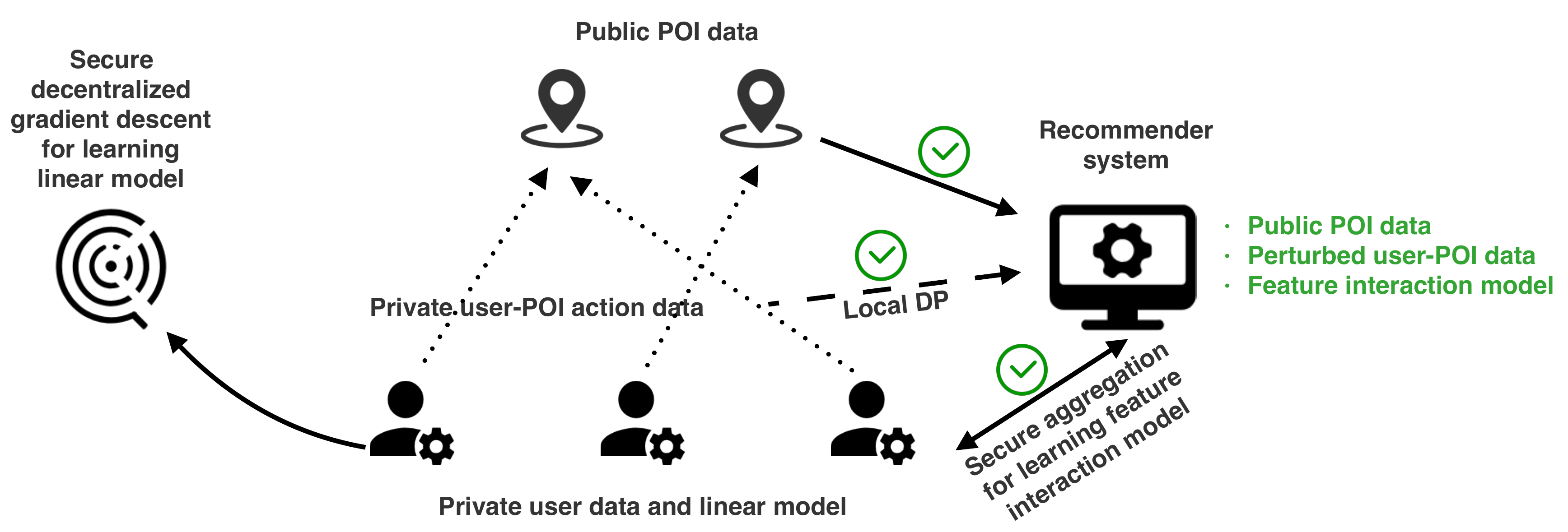}
%\vskip -0.1in
\caption{\texttt{Pri}vacy preserving POI \texttt{Rec}ommendation (\texttt{PriRec}) framework.}
\label{framework}
%\vskip -0.15in
\end{figure*}

\subsubsection{Motivation}
\label{sec-mot}
User privacy in POI recommendation should include two parts, i.e., the data that explicitly expose user privacy and the model that implicitly indicates user preferences or interests. 

\textbf{Data privacy. }
Both user and item (POI) features are important to recommendation performance. 
\textbf{User features} show the private information of users, e.g., age, occupation, and consumption ability, which are the most important information that need to be protected when building privacy preserving POI recommender system. 
One of the reasonable ways is to decentralize these private information on users' own device instead of collecting them. 
POI features show the static profile and dynamic operation status of the POI, both of which are public to all the users. 
The \textbf{POI static feature} are usually POI profiles, e.g., the dish category of a restaurant. 
The \textbf{POI dynamic feature} are usually operation status data, e.g., the check-in count of a hotel. 
However, these dynamic POI data are related to user-POI interaction histories, e.g., user-hotel check-in history, which are also a part of users' private data. 
Thus, techniques that can not only protect individual user-POI actions but also estimate the user-POI action count for each POI should be considered. 

To sum up, a privacy preserving POI recommender system should protect both user features and user-POI interaction data. 

\textbf{Model privacy. }
We take POI recommendation as a CTR prediction problem and design our model by following Factorization Machine (FM) \cite{rendle2012factorization}, since FM and its variants are popularly used due to its scalability and capability of capturing high-order feature interactions. 
Suppose each sample has $D$ real-valued features $\textbf{x} \in \mathbb{R}^{D}$, %a label $y$, and a 
its prediction $\hat{y}$ of the 2-order FM model is defined as, 
\begin{equation}\label{fm}
  \hat{y} = \underbrace{w_0 + \sum\limits_{d=1}^D{w_dx_d}}_{\text{linear model}} + \underbrace{\sum\limits_{d=1}^D\sum\limits_{d'=d+1}^D x_d x_{d'} \sum\limits_{k=1}^K v_{d,k} v_{d',k}}_{\text{feature interaction model}}.
\end{equation}

FM model has two parts, i.e., linear model and high-order feature interaction model. 
First, $w_0, w_1, ..., w_D$ are the linear model and each parameter $w_d$ denotes the weight of each feature $x_d$. 
Obviously, the linear model indicates the users' preferences on each feature and implicitly expose users' interests to some extent. 
Therefore, it should be kept privately by each user from being exposed to other users or the recommender. 
Second, $\textbf{V} \in \mathbb{R}^{D\times K}$ is the 2-order feature interaction model and $K$ is the dimensionality of feature interaction factorization. 
It can be seen that, %$w_{{d,d'}|d,d'=0,1,...,D}= 
$\sum_{k=1,...,K} v_{d,k} v_{d',k}$ is used to capture the weight of each feature interaction pair $<x_d,x_{d'}>$. 
Clearly, the weights of feature interaction pairs do not expose users' data or interests, and therefore, can be publish to the recommender. 

In summary, a privacy preserving POI recommender system should protect the sensitive models, e.g., the linear model of FM. 

\subsubsection{Notations and problem definition}

Formally, let $\mathcal{U}$ be the user set % with $I$ denoting user size, 
and $\textbf{X}^{i} \in \mathbb{R}^{m}$ be the private user features of user $i \in \mathcal{U}$. 
Let $\mathcal{V}$ be the item (POI) set %with $J$ denoting item size, 
and $\textbf{X}^{j} \in \mathbb{R}^{n}$ be the public POI feature of POI $j \in \mathcal{V}$. 
Let $(i,j)$ be an interaction between user $i \in \mathcal{U}$ and item $j \in \mathcal{V}$, $\textbf{X}^{ij} \in \mathbb{R}^{m+n} = \textbf{X}^{i} \textcircled{+} \textbf{X}^{j}$ be the feature\footnote{For simplification, we do not formalize contextual features such as distance and period of time.} 
of a sample with $\textcircled{+}$ be the concatenation operation, and $y_{ij} \in \{1,-1\}$ be the action, e.g., click or not. 
Let $\mathcal{O}$ be the training dataset, where all the user-item interactions $<\textbf{X}^{ij}, y_{ij}>$ are known. 

Let $\textbf{W}$ be the linear models of users with each row $\textbf{W}^i=w^i_0, w^i_1, ...,$ $w^i_{m+n}$ denotes the private linear model saved on the device of user $i \in \mathcal{U}$, and let $\textbf{V} \in \mathbb{R}^{(m+n)\times K}$ be the public feature interaction model hold by the recommender. 
The privacy preserving POI recommendation problem is to predict $\hat{y_{ij}}$ of unknow user-POI pairs, and meanwhile keeps $<\textbf{X}^{ij}, y_{ij}>$ and $\textbf{W}^i$ private. 
We summarize the notations used in this paper in Table \ref{notation}.

\begin{table}
\centering
\caption{Notation and description. }
\label{notation}
\begin{tabular}{|c|c|}
  \hline
  Notation & Description  \\
  \hline
  $\mathcal{U}$ & user set  \\
  \hline
  $\mathcal{V}$ & item set  \\
  \hline
  $\textbf{X}^{i}$ & private features of user $i \in \mathcal{U}$  \\
  \hline
  $\textbf{X}^{j}$ & public features of item $j \in \mathcal{V}$  \\
  \hline
  $(i,j)$ & an interaction between user $i \in \mathcal{U}$ and item $j \in \mathcal{V}$  \\
  \hline
  $\textbf{X}^{ij}$ & feature of an interaction $(i,j)$  \\
  \hline
  $y^{ij}$ & label of an interaction $(i,j)$  \\
  \hline
  $\hat{y_{ij}}$ & predicted label of an interaction $(i,j)$  \\
  \hline
  $\textbf{W}$ & private linear models  \\
  \hline
  $\textbf{W}^i$ & private linear model of user $i \in \mathcal{U}$  \\
  \hline
  $\textbf{V}$ & public feature interaction models  \\
  \hline
  $\lambda_w$ and $\lambda_v$ & regularization parameters  \\
  \hline
  $\nabla{w^i_d}$ & $d$-th element in the gradient of $\textbf{W}^i$ \\
  \hline
  $\nabla{\textbf{V}}$ & gradient of $\textbf{V}$ \\
  \hline
  $\alpha$ & learning rate  \\
  \hline
  $f \in \mathcal{N}(i)$ & neighbor of user $i \in \mathcal{U}$  \\
  \hline
  $S_{if}$ & relation strength between user $i$ and $f \in \mathcal{N}(i)$  \\
  \hline
  $K$ & factorization dimension of $\textbf{V}$ \\
  \hline
  $D$ & feature dimension  \\
  \hline
  $\sigma(x)$ & logistic function with input $x$  \\
  \hline
  $\langle a \rangle _j$ & $j$-th share of secret value $a$  \\
  \hline
  $\mathcal{A}_\epsilon$ & $\epsilon$-LDP randomized algorithm  \\
  \hline
  $\mathcal{O}$ & training dataset  \\
  \hline
\end{tabular}
\end{table}

\subsubsection{Model optimization}
In this paper, we take POI recommendation as a CTR prediction problem. 
The optimization task is to minimize the sum of losses $l$ over the training dataset $\mathcal{O}$
\begin{equation}
%OPT(\mathcal{O}) := 
\mathop {\arg \min }\limits_{\textbf{W},\textbf{V}} \sum\limits_{<\textbf{X}_{ij}, y_{ij}> \in \mathcal{O}} -ln (\sigma(y_{ij} \cdot \hat{y_{ij}})) + \lambda_w {||\textbf{W}||}_F^2 + \lambda_v {||\textbf{V}||}_F^2,
\end{equation}
where $\sigma(x) = 1/(1+e^{-x})$ is the logistic function, $\lambda_w$ and $\lambda_v$ are the regularization parameters for linear models and feature interaction model respectively, and $\hat{y_{ij}}$ is defined in Equation \eqref{fm}. 
For each user-POI pair $<\textbf{X}^{ij}, y_{ij}>$, its gradient with respect to each element $w^i_d$ in the linear model $\textbf{W}^i$ is 
\begin{equation}\label{linear-grad}
\nabla{w^i_d}=
\begin{cases}
y_{ij} \cdot (\sigma(y_{ij} \cdot \hat{y_{ij}}) - 1) + 2\lambda_w \cdot w^i_0,& \text{when $d=0$},\\
y_{ij} \cdot (\sigma(y_{ij} \cdot \hat{y_{ij}}) - 1) \cdot \textbf{X}^{ij}_d + 2\lambda_w \cdot w^i_d,& \text{when $d > 0$}.
\end{cases}
\end{equation}

Its gradient in terms of each element $V_{d,k}$ in the feature interaction model $\textbf{V}$ is 
\begin{equation}\label{inter-grad}
\nabla{V_{d,k}}= y_{ij} \cdot (\sigma(y_{ij} \cdot \hat{y_{ij}}) - 1) \cdot \textbf{X}^{ij}_d \sum\limits_{d' \ne d} V_{d',k} \textbf{X}^{ij}_{d'} + 2\lambda_v \cdot V_{d,k}.
\end{equation}

In traditional centralized setting, all the data and models are kept by the recommender, and FM can be efficiently learnt by using gradient descent \cite{rendle2012factorization}. 
In contrast, as we described earlier, in our privacy preserving setting, the private data and the linear models are decentralizedly hold by users. 
We will present how to learn the linear models and feature interaction model in the privacy preserving setting in Section \ref{section-learn-linear} and Section \ref{section-learn-inter}, respectively. 

\subsection{Overview of Privacy Preserving POI Recommendation Framework}
Our proposed \texttt{PriRec} framework can protect both private data and models, which is shown in Figure \ref{framework}. 
\textit{To protect data privacy}, users' private data, including features and actions, are decentralized on their own side, e.g., Cellphone or Pad. 
Besides, all the public POIs' data are kept by the recommender, and they are mainly in two types: the static data that describes the status of a POI such as POI category, and the dynamic data that indicate the popularity of a POI, e.g., visited count. 
%Since the user-POI actions are kept on users' devices, to obtain the statistics of these action data, we propose to adopt Local Differential Privacy (LDP) to generate dynamic POI features from the perturbed user-POI interaction data. 
\textit{To protect model privacy}, the linear models of \texttt{PriRec} are also decentralized on each user's side for privacy purpose, and we propose a \textit{secure decentralized gradient descent protocol} for users to learn them collaboratively. 
The feature interaction model is kept by the recommender, since it can only infer the interaction weights between features which has no privacy risk. 
We adopt \textit{secure aggregation strategy in federated learning} to learn it. 
To this end, both users' private data and models are kept by their own hands, and \texttt{PriRec} only collects the perturbed user-POI interaction data. 
Therefore, \texttt{PriRec} is able to protect both data and model privacy. 
We will present each part of the framework in details in the following sections. 

\subsection{Generating POI Dynamic Feature}

\begin{figure}[t]
\centering
\includegraphics[width=8cm]{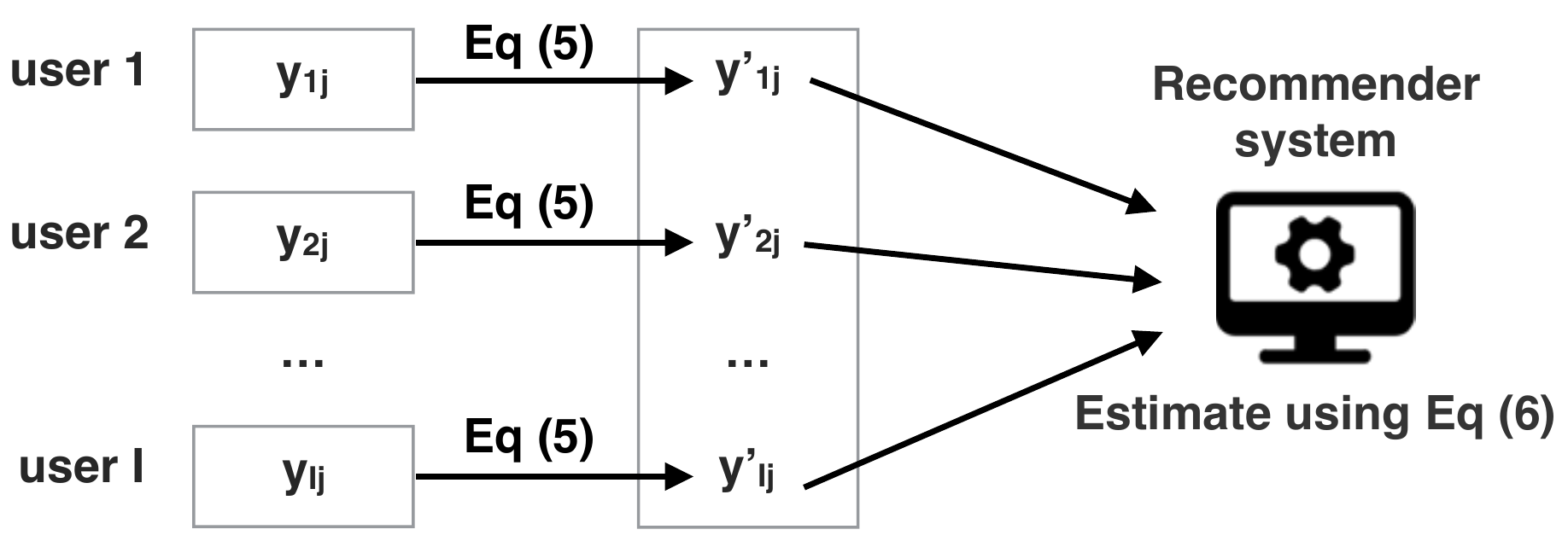}
%\vskip -0.1in
\caption{Generate dynamic features for POI $j$ using LDP.}
\label{ldp}
%\vskip -0.15in
\end{figure}

We propose to generate dynamic POI features, e.g., click count of a restaurant, by using Local Differential Privacy (LDP) to collect perturbed user-POI interaction data. 
In LDP, each user randomizes his/her private data using a randomized algorithm (mechanism) $\mathcal{A}$ locally, before sending them to data collector (recommender). 
\begin{definition}
 A randomized algorithm $\mathcal{A}_\epsilon$ : $\mathcal{V} \rightarrow \mathcal{Z}$ is $\epsilon$-locally differentially private ($\epsilon$-LDP) if for any pair of values $v, v' \in \mathcal{V}$ and any subset of output $\mathcal{S} \subset \mathcal{Z}$, we have that
  \begin{displaymath}
    \textbf{Pr}[\mathcal{A}(v) \in \mathcal{S}] \le e^\epsilon \cdot \textbf{Pr}[\mathcal{A}(v') \in \mathcal{S}].
  \end{displaymath}
\end{definition}
LDP formalizes a type of plausible deniability: no matter what output is released, it is approximately equally as likely to have come from one data point $v \in \mathcal{V}$ as any other \cite{bassily2015local,ding2017collecting}. 
In other words, the recommender can not differentiate whether a user has interaction with a POI or not, although it collects a perturbed user-POI interaction.
The user-POI interaction $y_{ij} \in \{0, 1\}$ is a binary value\footnote{We take $y_{ij}$ as in $\{0, 1\}$ when collecting data and as in $\{1,-1\}$ when learning model.}, which can be collected from users' devices by using the following mechanism:
\begin{equation}
y'_{ij} = 
\begin{cases}
1,& \text{with probability $\frac{1}{e^\epsilon + 1} + y_{ij} \cdot \frac{e^\epsilon - 1}{e^\epsilon +1 }$},\\
0,& \text{otherwise}.
\end{cases}
\end{equation}

After that, the recommender obtains the bits from all the users $\mathcal{U}$ and the total interaction count for POI $j$ can be estimated as 
\begin{equation}
c_{j} = \sum\limits_{i=1}^{I}\frac{y'_{ij} \cdot (e^\epsilon +1) - 1}{e^\epsilon - 1}.
\end{equation}

It can be proven that above data collection mechanism preserves $\epsilon$-LDP, and meanwhile achieves an unbiased estimation of the POIs' dynamic features \cite{ding2017collecting}. 
Besides dynamic visited count, LDP can also be used to estimate dynamic real-valued features, e.g., the average consumption of a POI. 
We finally show how to generate dynamic visited count features using LDP in Figure \ref{ldp}. 

% \begin{figure}[t]
% \centering
% \includegraphics[width=8cm]{figures/modelpro}
% \caption{Decentralized model propagation strategy.}
% \label{modelpro}
% \end{figure}

\subsection{Learning Linear Model}\label{section-learn-linear}
The linear models are decentralized on each users' devices for privacy concerns. 
Therefore, a key challenge is \textit{how should users collaboratively learn their linear models}. 
To solve this challenge, we first show the learning procedure of linear model in traditional centralized setting. 
By using gradient descent, the linear model is updated as follows
\begin{equation}\label{centgd}
{w_d}^{(t+1)} = {w_d}^{(t)} - \alpha \cdot \nabla{w_d}^{(t)},
\end{equation}
where $\alpha$ is the learning rate, and $\nabla{w_d}^{(t)}$ is the gradient of $w_d$ at time $t$. 
In decentralizing learning setting, data are hold by each individual learners and the traditional gradient descent is not suitable any more. 
Existing researches propose to approximate Equation \eqref{centgd} by using Decentralized Gradient Descent (DGD) \cite{nedic2009distributed,yuan2016convergence}, 
\begin{equation}\label{sdgd}
{w^i_d}^{(t+1)} = \sum\limits_{f \in \mathcal{N}(i)}S_{if} \cdot {w^f_d}^{(t)} - \alpha \cdot \nabla{w^i_d}^{(t)},
\end{equation}
where ${w^i_d}^{(t)}$ is the $d$-th model of user $i$ at time $t$, $\mathcal{N}(i)$ denotes the neighbors of $i$ on a certain user network, and $S_{if}$ denotes the edge weight between $i$ and $f$. 
We argue that DGD is not secure in our privacy preserving setting, since directly calculating the weighted sum of neighbors' linear models, i.e., $\sum_{f \in \mathcal{N}(i)}S_{if} \cdot {w^f_d}^{(t)}$, needs the plaintext model of neighbors, i.e., ${w^f_d}^{(t)}$, which directly reflects the preferences of users. 

\begin{algorithm*}[t]
\label{test}
\caption{Secure decentralized gradient descent protocol for learning the linear model of user $i$}
\KwIn {iteration $t$, linear model gradient of user $i$ ($\nabla{w^i_d}$), neighbors of user $i$ ($\mathcal{N}(i)$), linear models of neighbors (${w^f_d}^{(t)}, f \in \mathcal{N}(i) $), and the weight between $i$ and neighbors ($S_{if}, f \in \mathcal{N}(i)$)}
\For{each neighbor $f \in \mathcal{N}(i)$}{

	Calculates weighted linear model $sw^f_d = S_{if} \cdot {w^f_d}$ \\
	Locally generates shares $\left\langle sw^f_d \right\rangle_{j \in \mathcal{N}(i)}$ \\
	Keeps $\left\langle sw^f_d \right\rangle_{f}$ and distributes $\left\langle sw^f_d \right\rangle_{j \ne f}$ to neighbor ${j \in \mathcal{N}(i)}$

}
\For{each neighbor $f \in \mathcal{N}(i)$}{

	Locally calculates the summation of all $f$-th shares, i.e., $\sum\limits_{j \in \mathcal{N}(i)} \left\langle sw^j_d \right\rangle_{f}$ \\
	Sends $\sum\limits_{j \in \mathcal{N}(i)} \left\langle sw^j_d \right\rangle_{f}$ to user $i$ 
}
User $i$ calculates the summation from all neighbors, i.e., $\sum\limits_{f \in \mathcal{N}(i)}\sum\limits_{j \in \mathcal{N}(i)} \left\langle sw^j_d \right\rangle_{f}$, which equals to $\sum\limits_{f \in \mathcal{N}(i)}S_{if} \cdot {w^f_d}^{(t)}$ \\
User $i$ updates his/her linear model using Equation (\ref{sdgd}) \\
\end{algorithm*}

To solve the above problem, we propose a secure decentralized gradient descent protocol, as is shown in Algorithm 1. 
The main idea is to use secret sharing to calculate the summation of neighbors' linear models. 
Its security and correctness can be found in \cite{shamir1979share}. 
Note that the linear models are usually real-valued vectors, and we adopt the efficient fixed-point arithmetic method as described in Section 2.4, which has also been proven works in secret sharing settings. 
With the proposed secure decentralized gradient descent protocol, we can train the linear model without compromising users' private data and model. 

% We propose to solve this problem by using decentralized model propagation, 
% \begin{equation}\label{linear-update}
% {w^{(i \cup \mathcal{N}(i))}_d}^{(t+1)} = {w^{(i \cup \mathcal{N}(i))}_d}^{(t)} - \alpha \cdot S_{if} \cdot \nabla{w^i_d}^{(t)},
% \end{equation}
% where ${i \cup \mathcal{N}(i)}$ denotes the set of user $i$ and his/her neighbors. 
% We show our proposed decentralized model propagation in Figure \ref{modelpro}. 
% That is, when user $i$ has an interaction with POI $j$, $i$ calculates the linear model graident $\nabla{w^i_d}$ and updates the linear model $w^i_d$, meanwhile sends $\nabla{w^i_d}$ to neighbors to update their linear models. 

\begin{figure}[t]
\centering
\includegraphics[width=8cm]{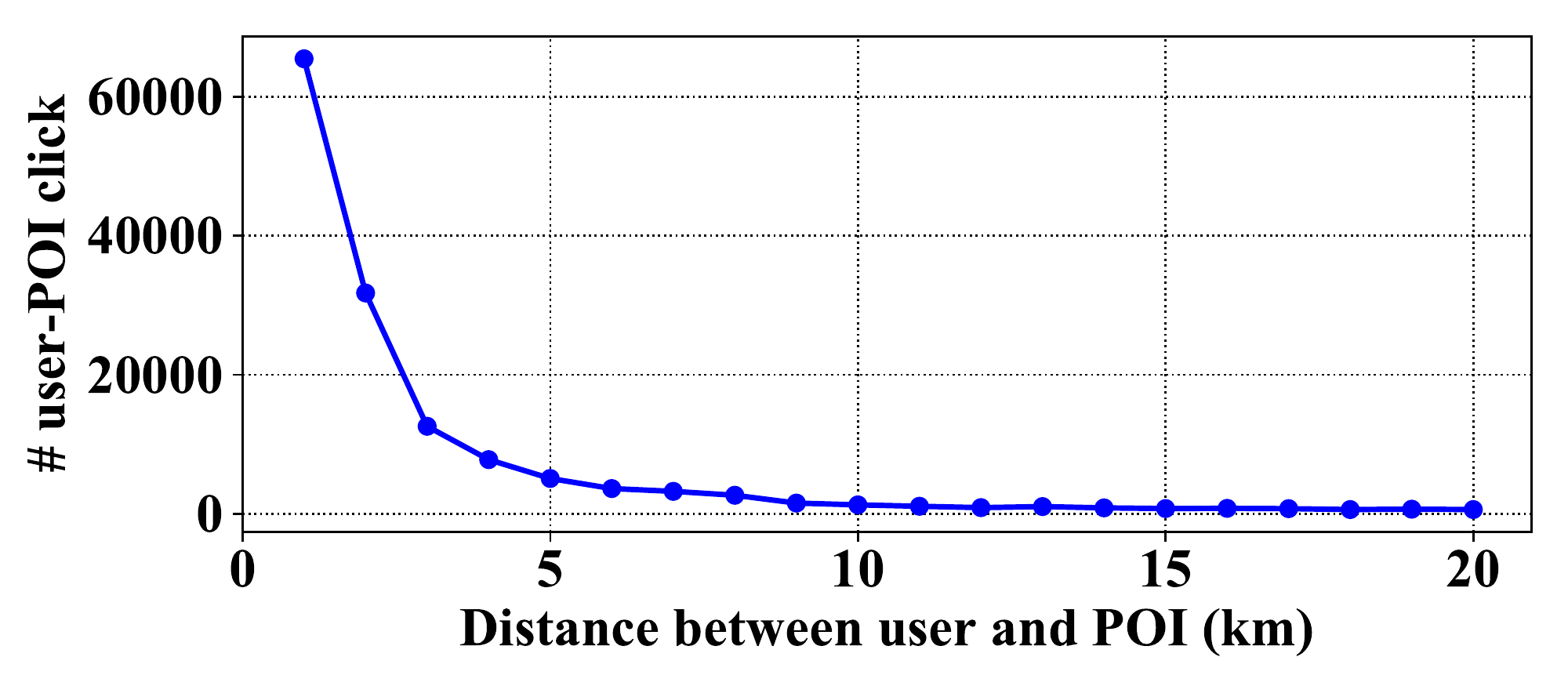}
%\vskip -0.1in
\caption{Relationship between user-POI distance and click.}
\label{distance}
%\vskip -0.15in
\end{figure}

The remaining challenge is \textit{how to choose neighbors for model propagation}. 
We address this challenge by analyzing the real data in POI recommendation channels from Koubei APP. 
Figure \ref{distance} shows the relationship between user-POI distances and actions. 
We can observe that, in practice, users tend to click the POIs nearby. 
In other words, POIs are likely to be interacted by the nearby users. 
Therefore, we build the user adjacent network by using user geographical information, similar as the existing researches \cite{ye2011exploiting,cheng2012fused}. 
Specifically, let $d_{i,f}$ be the distance between user $i$ and $f$, and the edge weight between $i$ and $f$ is defined as $S_{if}=f(d_{i,f})$, where $f(\cdot)$ is a mapping function that transforms distance to edge weight. 
Various mapping function has been proposed in literature \cite{zhao2016survey}. 

In practice, one can not communicate with all the other users, because (1) the communication cost is expensive, and (2) only a handful of users' devices are online. 
Therefore, for each user $i$, we randomly choose his/her closest top $N$ neighbors based on the distance. %and assume their devices are online and reachable. 
Further more, for simplification, we set the edge weights of the built user adjacent network to 1 after choosing neighbors, i.e., $S_{if}=1$. 
We will empirically study the effect of the number of maximum neighbors ($N$) on our model performance.

\subsection{Learning Feature Interaction Model}\label{section-learn-inter}
The feature interaction model is kept by the recommender, the relationship between the recommender and individual learners is similar as that of server and worker in parameter-server distributed learning paradigm \cite{li2014scaling}. 
The existing works propose secure aggregation strategy to train neural network model in federated learning settings \cite{konevcny2016federated,bonawitz2017practical}. 
Motivated by this, we adopt secure aggregation strategy in federated learning for users to learn the feature interaction model of FM collaboratively. 
Specifically, once a batch of online users have interactions with a POI, i.e., $<\textbf{X}^{ij}, y_{ij}>$, these users first pull the current feature interaction model $\textbf{V}^{(t)}$ from the recommender. 
They then calculate its gradient $\nabla{V}$ based on Equation \eqref{inter-grad}. 
After that, they securely aggragate the gradients $\nabla{V}$. 
Finally, the recommender updates the feature interaction model. 
% This secure aggregation strategy for learning the feature interaction model is summarized in Figure \ref{figsecagg}. 
In this paper, we adopt the most simple secure aggregation protocol in \cite{bonawitz2017practical}, i.e., one-time pad masking based on secret sharing. 
Please refer to Section 4.0.1 in \cite{bonawitz2017practical} for more details. 
Finally, the recommender updates $\textbf{V}^{(t)}$ as follows 
\begin{equation}\label{inter-update}
\textbf{V}^{(t+1)} = \textbf{V}^{(t)} - \alpha \cdot \nabla{\textbf{V}}^{(t)}.
\end{equation}

% \begin{figure}[t]
% \centering
% \includegraphics[width=9cm]{figures/secureagg}
% %\vskip -0.1in
% \caption{Secure aggregation for learning feature interaction model.}
% \label{figsecagg}
% %\vskip -0.15in
% \end{figure}

%Finally, the user pushs the updated feature interaction model ${V_{d,k}}^{(t+1)}$ to recommender. 
This strategy has the similar principle with the parameter server distributed learning paradigm \cite{li2014scaling}. 
That is, the server (i.e., the recommender) saves the model parameters ($\textbf{V}$), and the worker (i.e., each user) loads data and updates the models by communicating with the server. 
It becomes an asynchronous learning task when multi-users interact with POIs simultaneously, which is also a common task in parameter server \cite{li2014scaling}. 
%In this paper, we simply assume that there is only one user-POI interaction at each time point, and take the concurrent situation as a future work. 

\begin{algorithm}[t]\label{model-train}
\caption{\texttt{PriRec} Model Training}
\KwIn {training set ($\mathcal{O}$), learning rate ($\alpha$), regularization parameters ($\lambda_w$, $\lambda_v$), maximum propagation users ($N$), feature interaction factorization dimension ($K$), and maximum iterations ($T$)}
\KwOut{linear model for all the user ($\textbf{W}$) and \\ feature interaction model for the recommender ($\textbf{V}$)}

The recommender initializes $\textbf{V}$ \\
\For{Each user $i \in \mathcal{U}$}{
	Initialize $\textbf{W}^i$
}
\For{$t=1$ to $T$}
{
	Shuffle training data $\mathcal{O}$\\
	\For{each user-POI pair $<\textbf{X}^{ij}, y_{ij}> \in \mathcal{O}$, user $i$}{
		
		\# \textbf{learn linear model} \\
		Calculate $\nabla{\textbf{W}^i}$ based on Equation \eqref{linear-grad} \\
		Update $\textbf{W}^i$ based on the secure decentralized gradient descent protocol in Algorithm 1 \\
		% \For{user $f$ in top $N$ of $\mathcal{N}(i)$}{
		% 	Receive $\nabla{\textbf{W}^i}$ from user $i$ \\
		% 	Update $\textbf{W}^f$ based on Equation \eqref{linear-update}
		% }

		\# \textbf{learn feature interaction model} \\
		Pull $\textbf{V}$ from the recommender \\
		Calculate $\nabla{\textbf{V}}$ based on Equation \eqref{inter-grad} \\
		Push $\nabla{\textbf{V}}$ to the recommender using secure aggregation \\
		The recommender updates $\textbf{V}$ based on Equation \eqref{inter-update} \\
	}
}
\Return $\textbf{W}$ and $\textbf{V}$
\end{algorithm}

\subsection{Model Training and Prediction Algorithm}
The training of \texttt{PriRec} includes two parts, i.e., learning linear models and learning feature interaction model, and we summarize it in Algorithm 1. 
As we have described in Section \ref{section-learn-linear}, linear models are decentralized on each users' devices for privacy concerns, and we propose a \textit{secure decentralized gradient descent protocol} for users to learn them collaboratively. 
We summarize the learning algorithm in \texttt{lines} 9-10. 
The feature interaction model is kept by the recommender, and we adopt \textit{secure aggregation strategy} in federated learning for users to learn collaboratively, as is presented in Section \ref{section-learn-inter}, which corresponds to \texttt{lines} 12-15 in Algorithm 2. 

The prediction of \texttt{PriRec} also needs the communication between users and the recommender, as is shown in Algorithm 3. 
In it, \texttt{line} 1 denotes the matching procedure before ranking. 
Different matching strategies can be used, e.g., the simplest location based matching strategy. 
We do not describe the matching strategies in details, because it is not the focus of this paper. 
Line 5 omits the contextual features for conciseness. 

In summary, \texttt{PriRec} is able to protect users' private data and model during model training and prediction procedures. 
Similar to most prior privacy preserving machine learning algorithms \cite{mohassel2017secureml}, \texttt{PriRec} can only protect against \textit{semi-honest} adversary using secret sharing technique. That is, \texttt{PriRec} assumes the participants strictly follow the protocol execution. 
%The security proof can be refered to \cite{shamir1979share}, since we apply secret sharing technique. 
We leave how to solve \textit{malicious} adversary as a future work.

\begin{algorithm}[t]\label{model-predict}
\caption{\texttt{PriRec} Model Prediction for User $i$}
\KwIn {features for user $i$ ($\textbf{X}^{i}$), features for each POI $j$ ($\textbf{X}^{j}$), linear model for user $i$ ($\textbf{W}^i$) on his/her device, and feature interaction model on server ($\textbf{V}$)}
\KwOut{recommend top $k$ POIs $\mathcal{V}_k$ for user $i$}

Get the matched POI set $\mathcal{V}_m \in \mathcal{V}$ \\
Pull $\textbf{V}$ from the recommender \\
\For{each POI $j \in \mathcal{V}_m$, user $i$}{

	\# \textbf{combine features} \\
		Pull POI $j$'s feature $\textbf{X}^{j}$ from the recommender \\
		Concat user $i$'s feature $\textbf{X}^{i}$ with POI $j$'s feature $\textbf{X}^{j}$, and get $\textbf{X}^{ij}$

	\# \textbf{predict score} \\
	Predict user $i$'s score on POI $j$ based on Equation \eqref{fm} \\
}

Recommend top $k$ POIs $\mathcal{V}_k$ for user $i$ with the highest scores \\
\Return $\mathcal{V}_k$ for user $i$
\end{algorithm}

\subsection{Complexity Analysis}\label{sec-complexity}
We now analyze the communication and computation complexities of Algorithm 1. 
Recall that $|\mathcal{O}|$ is the training data size, $D$ is the feature size, $K$ is the dimensionality of feature interaction factorization, and $N$ denotes the number of maximum neighbors to be communicated. 

\noindent
\textbf{Communication Complexity.} 
For each user-POI pair, the communication relies on two parts. 
(1) users communicate with each other to learn linear models, i.e., \texttt{lines} 9-10, and its complexity is $O(N^2 \cdot D)$; 
(2) users communicate with the recommender to learn feature interaction model, i.e., \texttt{lines} 12-15, and its complexity is $O(2 D \cdot K)$. 
Therefore, the total communication cost in Algorithm 1 is $O(|\mathcal{O}| \cdot D \cdot (N^2 + 2K))$. 
In practice, since $D, N, K \ll |\mathcal{O}|$, the total communication complexity is linear with data size. 

\noindent
\textbf{Computation Complexity.}
For each user-POI pair, the computing bottleneck is Equation \eqref{fm}, and it has linear computation complexity $O(K \cdot D)$ after reformulating it \cite{rendle2012factorization}. 
Therefore, the computation complexity of learning linear models, i.e., \texttt{lines} 9-10, is $O(N \cdot D)$; 
The computation complexity of learning feature interaction model, i.e., \texttt{lines} 12-15, is $O(K \cdot D)$. 
In total, the computation cost in Algorithm 1 is also $O(|\mathcal{O}| \cdot D \cdot (N + 2K))$. 
Since $D, N, K \ll |\mathcal{O}|$, the total computation complexity is also linear with data size. 

For Algorithm 2, we analyze that, for predicting each user-POI pair, the communication complexity is $O(|\textbf{X}^{j}| + K \cdot D)$ and the computation complexity is $O(K \cdot D)$, where $|\textbf{X}^{j}|$ denotes the feature size of POI $j$. 
In practice $K$ is usually very small, therefore, the complexities are linear with $D$.

%\subsection{Discussion} model compression 

\section{Empirical Study}	
In this section, we empirically compare the performance of the proposed \texttt{PriRec} with the existing non-private POI recommendation method. 
We also study the effects of parameters on model performance.

\subsection{Setting}

\begin{table}[h]
\centering
\caption{Dataset description}
%\vskip -0.15in
\label{dataset}
\begin{tabular}{|c|c|c|c|c|}
  \hline
  Dataset & \#user & \#item & \#interaction & \#feature  \\
  \hline
  \emph{Foursquare} & 11,824 & 13,924 & 924,474 & 6 \\
  \hline
  \emph{Koubei} & 85,466 & 118,598 & 497,838 & 89 \\
  \hline
\end{tabular}
%\vskip -0.15in
\end{table}

\textbf{Datasets.} We choose two real-world user-POI interaction datasets for experiments, i.e., \emph{Foursquare} and \emph{Koubei}. 

First, \emph{Foursquare} a famous benchmark dataset for POI recommendation \cite{yang2016privcheck}. 
It contains user-POI action histories in two cities, and we only choose the data in Tokyo. 
We filter the POIs which are interacted by less than 10 users. 
Since \emph{Foursquare} only has positive user-POI interaction data, we randomly sample 1 negative user-POI interactions for each record, and therefore, the ratio of positive and negtive records is 1:1. 
The original dataset only has \textbf{user features} such as gender, friend count. We also generate \textbf{POI dynamic features} using our proposed local DP technique. 
Moreover, since the \emph{Foursquare} dataset does not have the geographic locations when a user interacts with a POI, we can not build user geographic adjacent network. 
Instead, we build the user adjacent network by random. That is, we randomly select ($N$) neighbors for each user-POI interaction. 

Second, the \emph{Koubei} dataset is collected from the POI recommendation channel in \emph{Koubei}\footnote{https://www.koubei.com/}, which is a product of Alibaba and Ant Financial in China, and we filter the users and POIs whose interactions are less than 5. 
There are many kinds of POIs in this channel, such as restaurants, cinema, and markets. 
The \emph{Koubei} dataset consists of two parts, the positive user-POI interaction ($y_{ij}=1$) indicates that a user clicks on a POI, the negative user-POI interaction ($y_{ij}=0$) implies that a user ignores a POI after exposure, and their ratio is about 1:3. 
The \emph{Koubei} dataset has three kinds of features, as described in Section \ref{sec-mot}, i.e., \textbf{user features} such as hometown and gender, \textbf{POI static features} like POI category, and the generated \textbf{POI dynamic features} such as recently clicked count of POIs. 
The \emph{Koubei} dataset has geographic location information, with which we build user geographic adjacent network, and we use $N$ to denote the maximum number of neighbors for each user. 

Finally, Table \ref{dataset} shows the statistics of both datasets after pre-process. 

\textbf{Metrics.} 
Since we take POI recommendation as a CTR prediction problem in this paper, we adopt Area Under the receiver operating characteristic Curve (AUC) as the evaluation metric, which is commonly used to evaluate CTR prediction quality \cite{ling2017model}. 
In practice, AUC of a classifier is equivalent to the probability that the classifier will rank a randomly chosen positive instance higher than a randomly chosen negative instance \cite{fawcett2006introduction}, therefore, the higher the better. 

We split both datasets with two strategies: (1) randomly sample 80\% as training set and the rest 20\% as test set, and (2) randomly sample 90\% as training set and the rest 10\% as test set.
We use \emph{Foursquare80} and \emph{Koubei80} to denote the first strategy, and use \emph{Foursquare90} and \emph{Koubei90} to denote the second strategy.
We repeat this procedure three times and report their average results. 

% \begin{table*}
% \centering
% \caption{Summary of the existing models and our proposed models, where ? FM means we want to study the corresponding model performance with FM during experiments. }
% \label{compare-summary}
% \begin{tabular}{|c|c|c|c|}
%   \hline
%   Model & Data & Information Leakage & Performance   \\
%   \hline
%   \textbf{MF} & rating & data & -  \\
%   \hline
%   \textbf{DMF} & rating & gradient & $\approx$ \textbf{MF} \cite{chen2018privacy} \\
%   \hline
%   \textbf{FM} & rating, user feature, POI static feature, POI dynamic feature & data & > \textbf{MF} \cite{rendle2012factorization} \\
%   \hline
%   \textbf{\texttt{PriRec-}} & rating, user feature, POI static feature & no information leadage & ? \textbf{FM} \\
%   \hline
%   \textbf{\texttt{PriRec}} & rating, user feature, POI static feature, POI dynamic feature & no information leadage & ? \textbf{FM} \\
%   \hline
% \end{tabular}
% \end{table*}

\begin{table}[th]
\centering
\caption{Summary of the existing models and our proposed models, where users' private information are shown in italics and ?FM means we want to study the corresponding model performance with FM during experiments. }
\label{compare-summary}
\begin{tabular}{|c|c|c|c|}
  \hline
  Model & Data & \tabincell{c}{Information \\Leakage} & Performance   \\
  \hline
  \textbf{MF} & \textit{rating} & \textit{rating} & -  \\
  \hline
  \textbf{DMF} & \textit{rating} & \textit{gradient} & $\approx$ \textbf{MF} \cite{chen2018privacy} \\
  \hline
  \textbf{FM} & \tabincell{c}{\textit{rating} \\ \textit{user feature}  \\ POI static feature \\ POI dynamic feature} & \tabincell{c}{\textit{rating} \\ and \\ \textit{user feature}} & > \textbf{MF} \cite{rendle2012factorization} \\
  \hline
  \textbf{\texttt{PriRec-}} & \tabincell{c}{\textit{rating} \\ \textit{user feature} \\POI static feature} & \tabincell{c}{no \\ private\\information \\ leakage} & ?\textbf{FM} \\
  \hline
  \textbf{\texttt{PriRec}} & \tabincell{c}{\textit{rating} \\ \textit{user feature} \\ POI static feature \\ POI dynamic feature} & \tabincell{c}{no \\ private\\information \\ leakage} & ?\textbf{FM} \\
  \hline
\end{tabular}
\end{table}

\begin{table*}[t]
\centering
\caption{AUC comparison on \emph{Foursquare} datasets }
%\vskip -0.1in
\label{compare-fs}
\begin{tabular}{|c|c|c|c|c|c|c|c|}
 \hline
  Datasets & \multicolumn{3}{c|}{\emph{Foursquare80}} & \multicolumn{3}{c|}{\emph{Foursquare90}}  \\
  \hline
  \hline
  Model & FM & \texttt{PriRec-} & \texttt{PriRec} & FM & \texttt{PriRec-} & \texttt{PriRec} \\
  \hline
  $K=5$ & \textbf{0.8152} & 0.4777 & 0.7834 & \textbf{0.8106} & 0.4722 & 0.7818 \\
  \hline
  $K=10$ & \textbf{0.8145} & 0.4771 & 0.7831 & \textbf{0.8098} & 0.4727 & 0.7824 \\
  \hline
  $K=15$ & \textbf{0.8131} & 0.4749 & 0.7829 & \textbf{0.8083} & 0.4702 & 0.7816 \\
  \hline
\end{tabular}
\end{table*}

\begin{table*}[t]
\centering
\caption{AUC comparison on \emph{Koubei} datasets }
%\vskip -0.1in
\label{compare-kb}
\begin{tabular}{|c|c|c|c|c|c|c|c|}
 \hline
  Datasets & \multicolumn{3}{c|}{\emph{Koubei80}} & \multicolumn{3}{c|}{\emph{Koubei90}}  \\
  \hline
  \hline
  Model & FM & \texttt{PriRec-} & \texttt{PriRec} & FM & \texttt{PriRec-} & \texttt{PriRec} \\
  \hline
  $K=5$ & 0.7154 & \textbf{0.7484} & \textbf{0.7605} & 0.7172 & \textbf{0.7495} & \textbf{0.7695} \\
  \hline
  $K=10$ & 0.7180 & \textbf{0.7519} & \textbf{0.7633} & 0.7205 & \textbf{0.7534} & \textbf{0.7713} \\
  \hline
  $K=15$ & 0.7192 & \textbf{0.7529} & \textbf{0.7643} & 0.7207 & \textbf{0.7546} & \textbf{0.7720} \\
  \hline
\end{tabular}
\end{table*}

\textbf{Comparison methods.} 
Our proposed \texttt{PriRec} framework is a novel decentralized algorithm of the existing Factorization Model (FM) \cite{rendle2012factorization}, and it belongs to privacy-preserving decentralized recommendation approaches. 
FM has been proven outperform the existing Matrix Factorization (MF) \cite{mnih2007probabilistic} model due to its ability to handle additional feature information besides the user-item interaction (rating) information. 
As long as the features are useful, which is always so in practice, FM can beat MF consistantly. 
%Similarly, the existing Decentralized Matrix Factorization (DMF) \cite{chen2018privacy} model can only handle rating information, which is not fair to compare with. 
Therefore, we only compare our proposed model with FM. 
Moreover, we would like to study the contribution of the generated dynamic POI features to the accuracy of \texttt{PriRec}, and therefore, we use \texttt{PriRec-} to indicate the version that \texttt{PriRec} does not use the generated dynamic POI features by LDP. 
We summarize the characteristics of the above mentioned models in Table \ref{compare-summary}. 
From it, we can see that our proposed \texttt{PriRec} framework can utilize more information without compromising users' private data. 

% We compare \texttt{PriRec} with the following classic centralized recommendation approaches, and the state-of-the-art privacy-preserving decentralized recommendation approach: 
% factor models,
% \begin{itemize}[leftmargin=*] \setlength{\itemsep}{-\itemsep}
%     \item \textbf{Matrix Factorization (MF)} \cite{mnih2007probabilistic} is a classic centralized collaborative filtering model, and it factorizes the user-item rating matrix into real-valued user and item latent factor matrixes.
%     \item \textbf{FM} \cite{rendle2010factorization} is a classic centralized CTR prediction model, and it improves logistic regression by considering high-order feature interactions. 
%     \item \textbf{Decentralized Matrix Factorization (DMF)} \cite{chen2018privacy} is the state-of-the-art privacy-preserving decentralized recommendation approach, and it can only handle rating information.
% \end{itemize}

\textbf{Hyper-parameters.} 
% We set learning rate $\theta=0.001$, and regularization parameters $\lambda_w=\lambda_v=0.001$. 
%For the latent factor dimension of MF and DMF, and the feature interaction factorization dimension of FM and prirec, we change their values in $\{5,10,15,20\}$. 
We set $\epsilon=1$ for LDP when generating dynamic POI features, following the existing research \cite{ding2017collecting}. 
We vary the number of maximum neighbors ($N$) and the feature interaction factorization dimension ($K$) of FM and \texttt{PriRec} to study their effects on model performance, and vary the maximum number of iterations ($T$) to study its effect on model convergency. 
We find the best values of other hyper-parameters, including learning rate ($\alpha$) and regularization parameters ($\lambda_w$ and $\lambda_v$), in $\{10^{-4},10^{-3},10^{-2},10^{-1},10^{0}\}$.

\subsection{Comparison Results}
We compare \texttt{PriRec} and \texttt{PriRec-} with the classic FM model on both \emph{Foursquare} and \emph{Koubei} datasets. 
Note that during the comparison, we use grid search to find the best parameters of each model. 

\noindent
\textbf{Results on \emph{Foursquare}.} 
We first report the comparison results on \emph{Foursquare} in Table \ref{compare-fs}. From it, we find that 
\begin{itemize}%[leftmargin=*] \setlength{\itemsep}{-\itemsep}
    \item In most of the cases, the recommendation performance of each model decreases with training data size and $K$, where $K$ is the dimensionality of feature interaction factorization. This is because the \emph{Foursquare} dataset only has 6 features, including 3 dynamic POI features generated by LDP, which causes over-fitting problem. 
    \item The AUC performance of \texttt{PriRec-} is even less than 0.5 (random guss), which is quite unsatisfying. This is because the original \emph{Foursquare} dataset only has 3 user features, with which it is unable to train a reasonable model. 
    \item Our proposed dynamic POI popularility features generated by using LDP can significantly improve the recommendation performance of \texttt{PriRec}. 
    For example, the AUC of \texttt{PriRec} improves 65.57\% comparing with that of \texttt{PriRec-} on \emph{Foursquare90} when $K=5$. 
    \item \texttt{PriRec} and FM have comparable recommendation performance (0.78+ vs. 0.81+). That is, our proposed model can protect user privacy by sacrificing little recommendation accuracy. 
\end{itemize}

\noindent
\textbf{Results on \emph{Koubei}.} 
We then report the comparison results on \emph{Koubei} in Table \ref{compare-kb}. 
We observe that:
\begin{itemize}%[leftmargin=*] \setlength{\itemsep}{-\itemsep}
    %\item Recommendation performance of each model increases with training data size. For example, the AUC of \texttt{PriRec} on \emph{Koubei90} improves 7.59\% comparing with that on \emph{Koubei80} when $K=5$. This is because the model is better trained, given more training data. 
    \item Recommendation performance of each model increases with $K$. $K$ is the dimensionality of feature interaction factorization, and therefore, with enough features, the bigger $K$ is, the better the learnt feature interaction model $\textbf{V}$ captures the real relations between features. 
    \item Our proposed dynamic POI popularility features generated by using LDP can significantly improve the recommendation performance of \texttt{PriRec}. 
    For example, the AUC of \texttt{PriRec} improves 2.67\% comparing with that of \texttt{PriRec-} on \emph{Koubei90} when $K=5$. 
    \item Recommendation performance of \textbf{\texttt{PriRec} consistently outperforms FM in all the cases}. 
    For example, the AUC of \texttt{PriRec} improves that of FM as high as 6.30\% on \emph{Koubei80} when $K=5$. 
    Note that, FM uses all the features, including the dynamic POI features, in the traditional centralized training setting. 
    \textbf{The reason is}, in POI recommendation scenarios, the user-POI interactions obey \textit{location aggregation}, i.e., most users only active in a certain location. Different from FM, which has a centralized linear model, \texttt{PriRec} belongs to decentralized model and it learns the linear models for different users by using secure decentralized gradient descent. To this end, \texttt{PriRec} is able to capture users' individual interests in different locations. 
    This is consistent with the reality that users in different places have different tastes. 
\end{itemize}

\begin{figure*}[t]
\centering
\subfigure [Average training loss on \emph{Foursquare80}]{ \includegraphics[width=5cm]{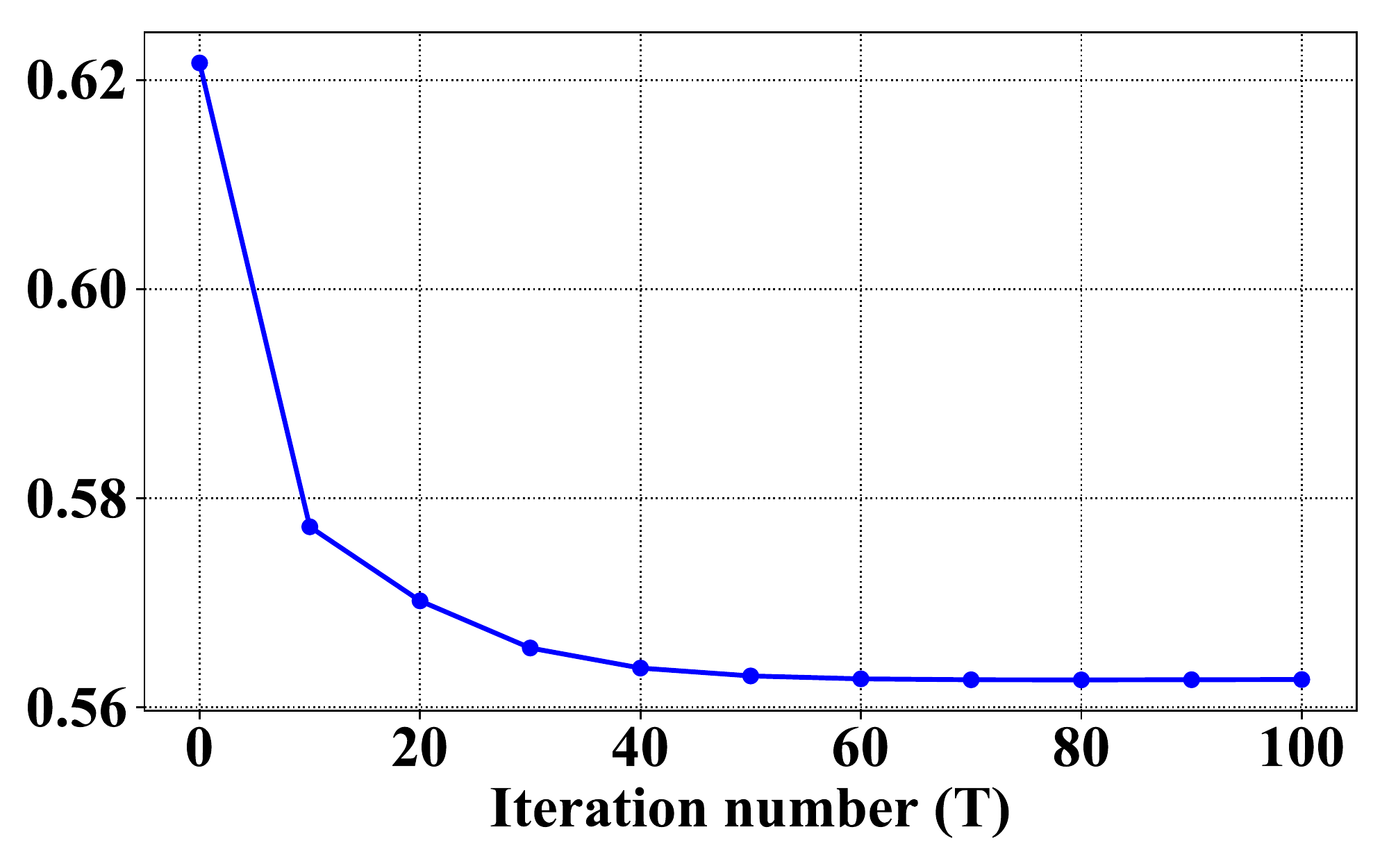}}
\subfigure [Average test loss on \emph{Foursquare80}] { \includegraphics[width=5cm]{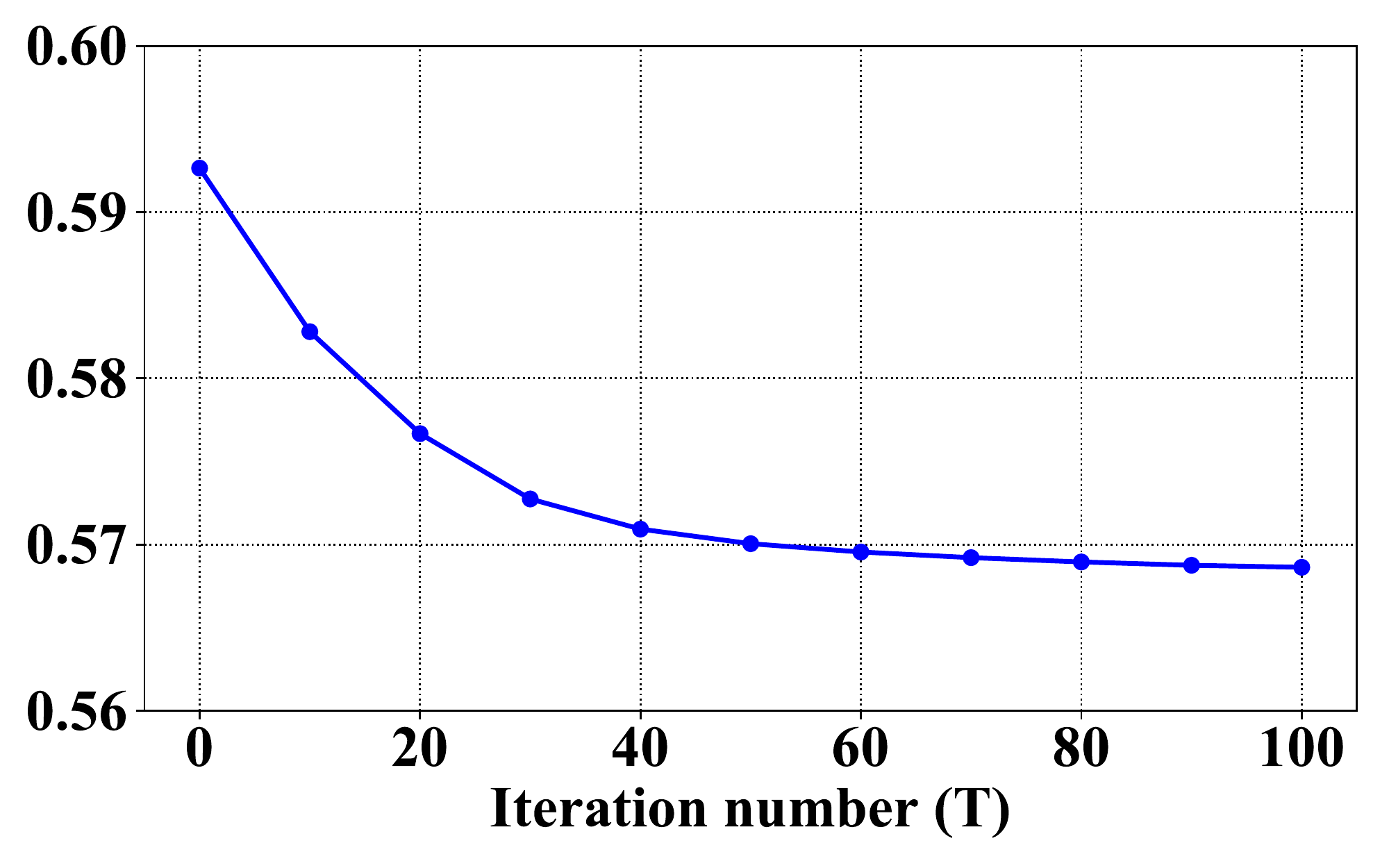}}
\subfigure [Average training loss on \emph{Koubei80}]{ \includegraphics[width=5cm]{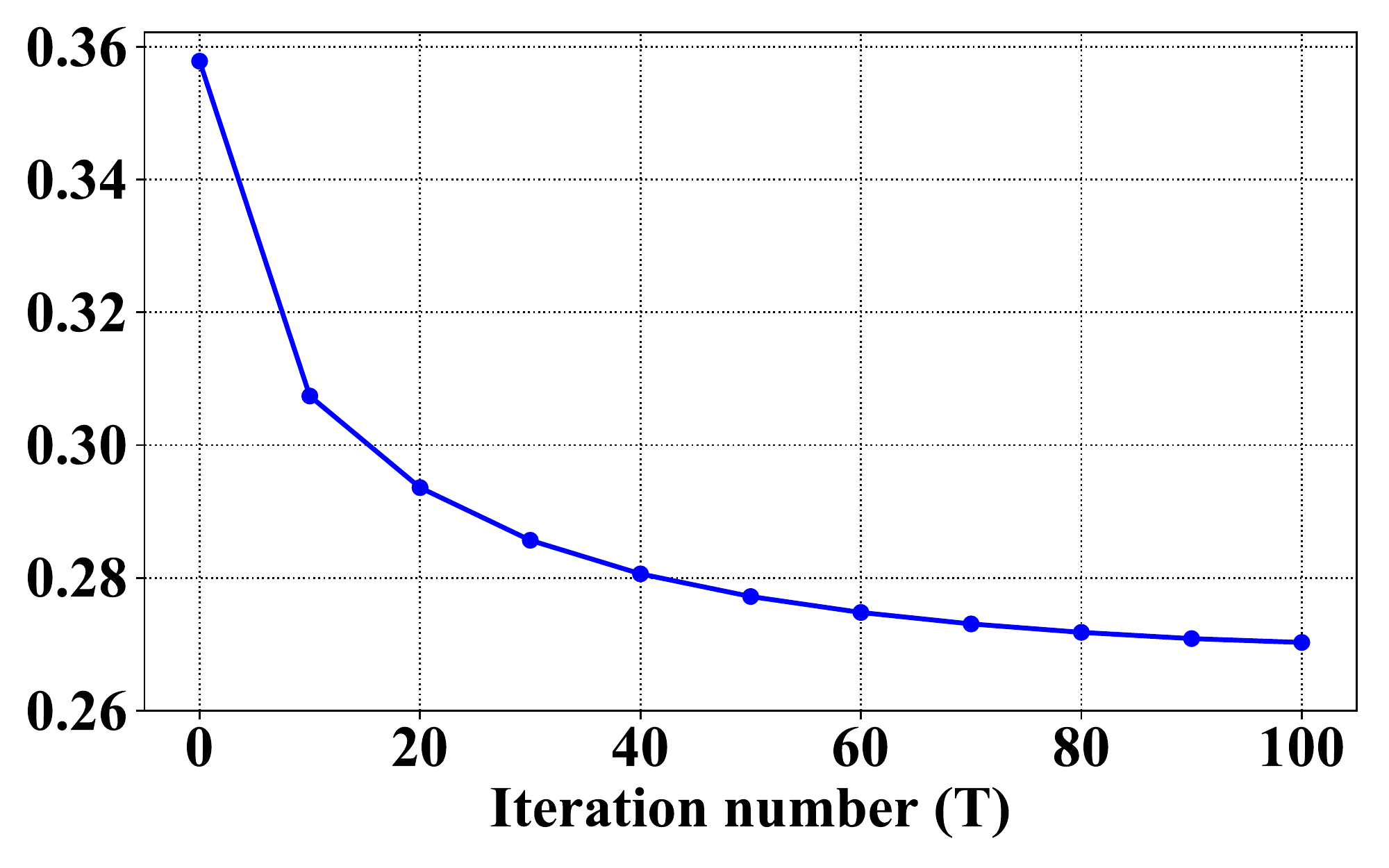}}
\subfigure [Average test loss on \emph{Koubei80}] { \includegraphics[width=5cm]{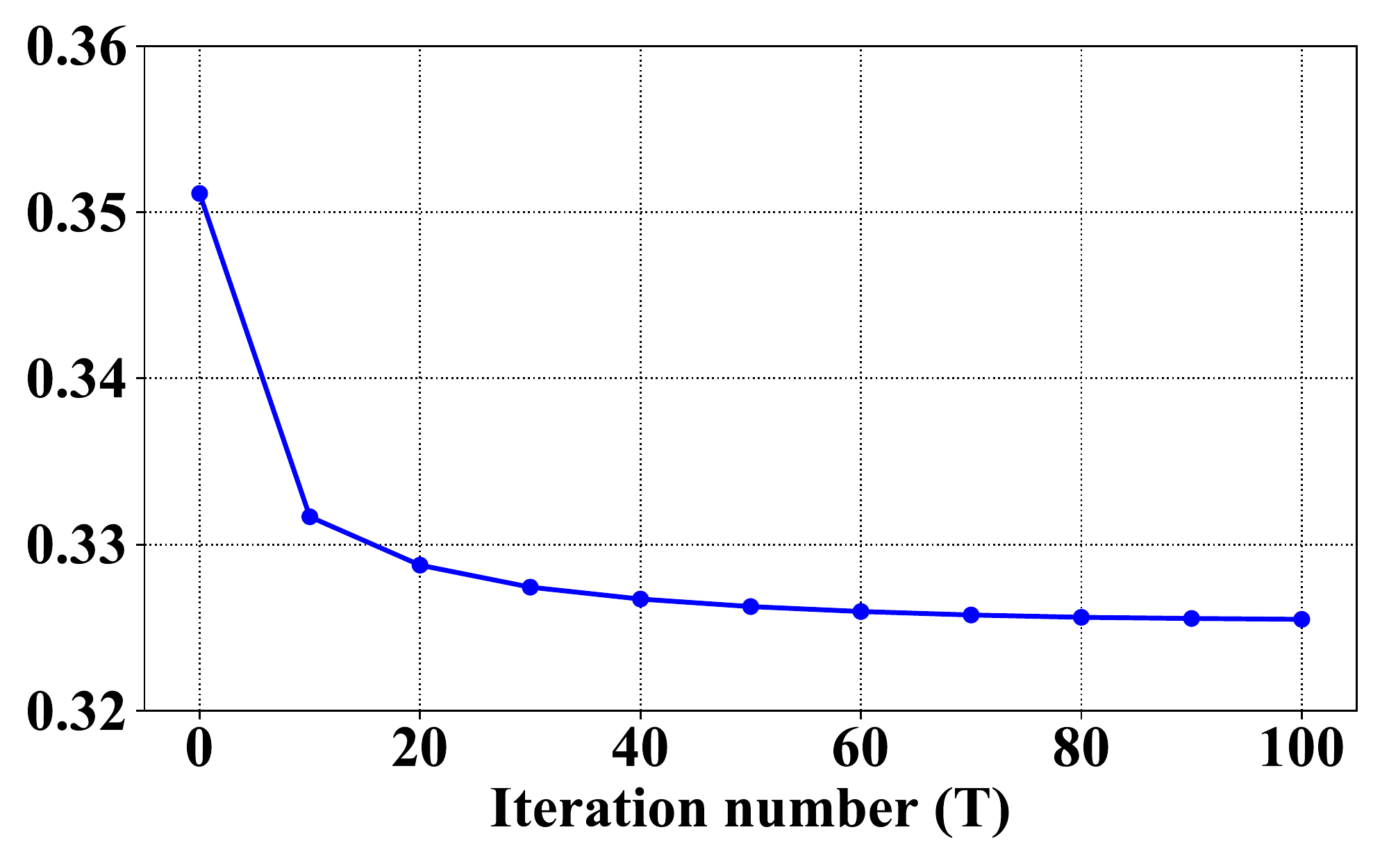}}
%\vskip -0.15in
\caption{Average trainning and test losses of \texttt{PriRec} w.r.t. the number of iteration number ($T$).}
\label{effectt}
\end{figure*}

\subsection{Parameter Analysis}
%Local DP $\epsilon$
We first analyze the convergence of \texttt{PriRec} in this section.
%We first analysis the effect of maximum iteration ($T$) on model convergence. 
We show the average training loss and test loss of \texttt{PriRec} w.r.t. the number of iteration number ($T$) in Figure \ref{effectt}, where we set $K=5$ and the number of maximum neighbors $N=30$. 
It obviously shows that \texttt{PriRec} converges faster on \emph{Foursquare80} than \emph{Koubei80}. 
This is because there are only 6 features on \emph{Foursquare} dataset, in contrast, there are 89 features on \emph{Koubei}.

\begin{figure*}[t]
\centering
\subfigure [\texttt{PriRec-} on \emph{Foursquare80}]{ \includegraphics[width=5cm]{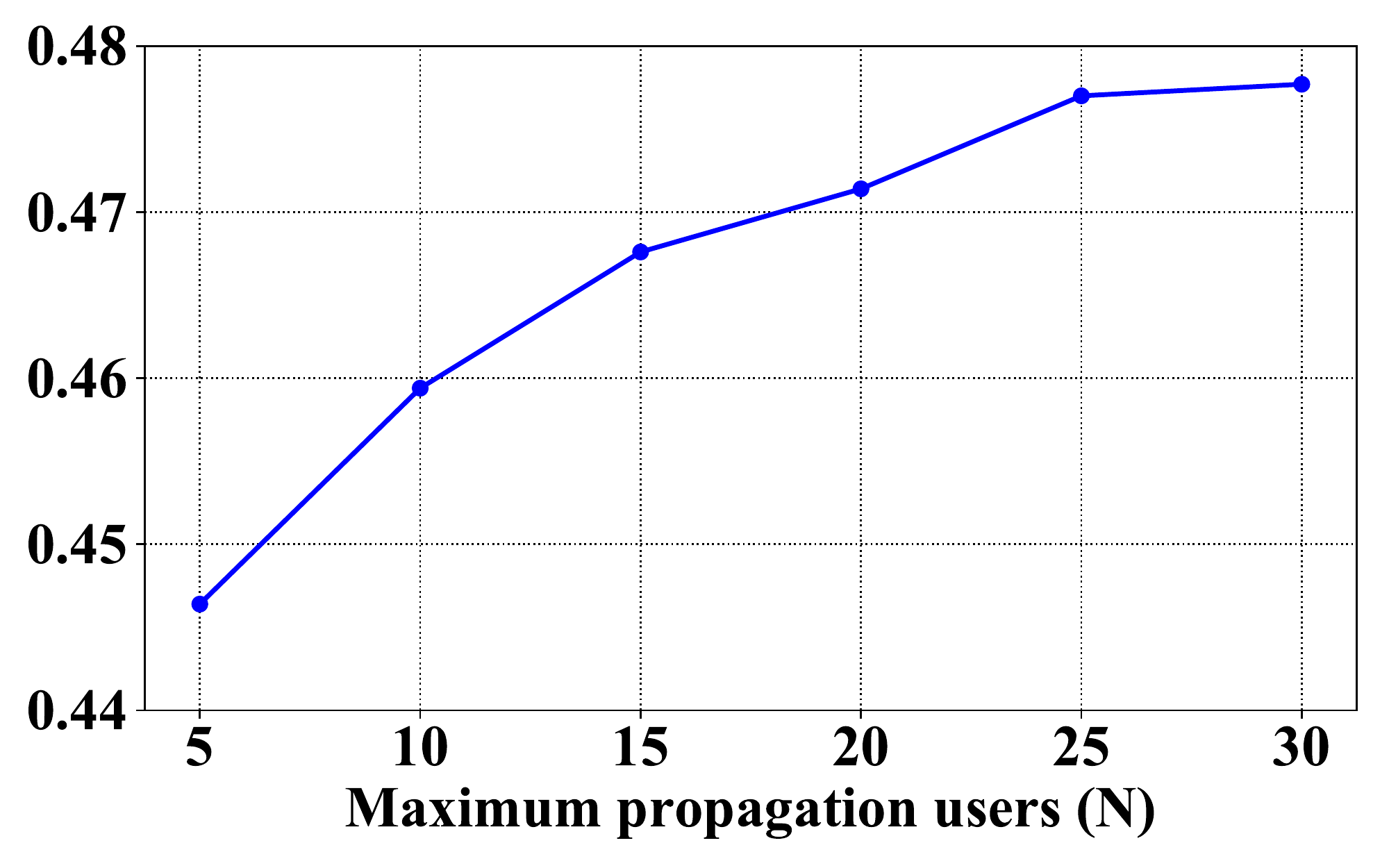}}
\subfigure [\texttt{PriRec-} on \emph{Foursquare90}] { \includegraphics[width=5cm]{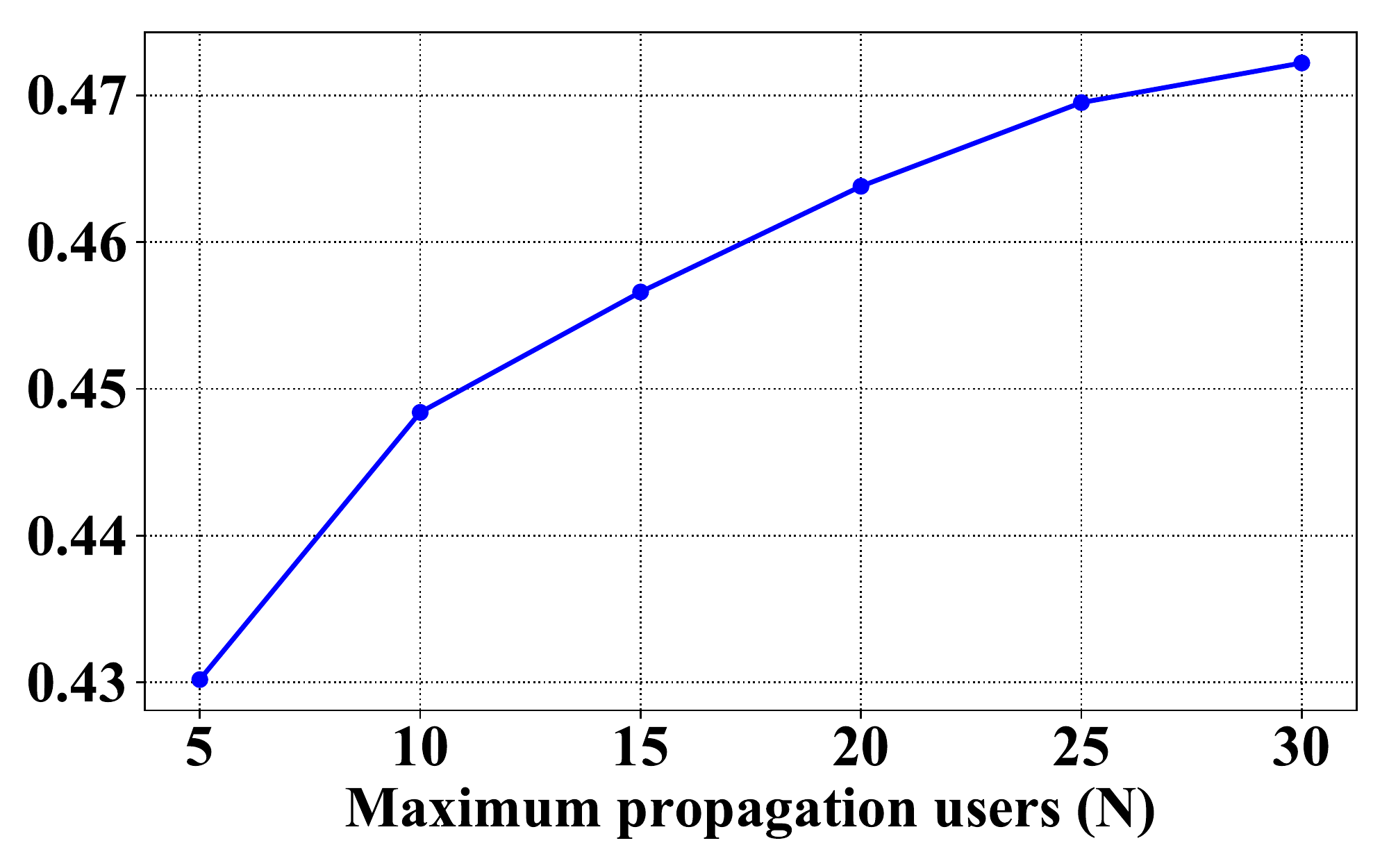}}
\subfigure [\texttt{PriRec} on \emph{Koubei80}]{ \includegraphics[width=5cm]{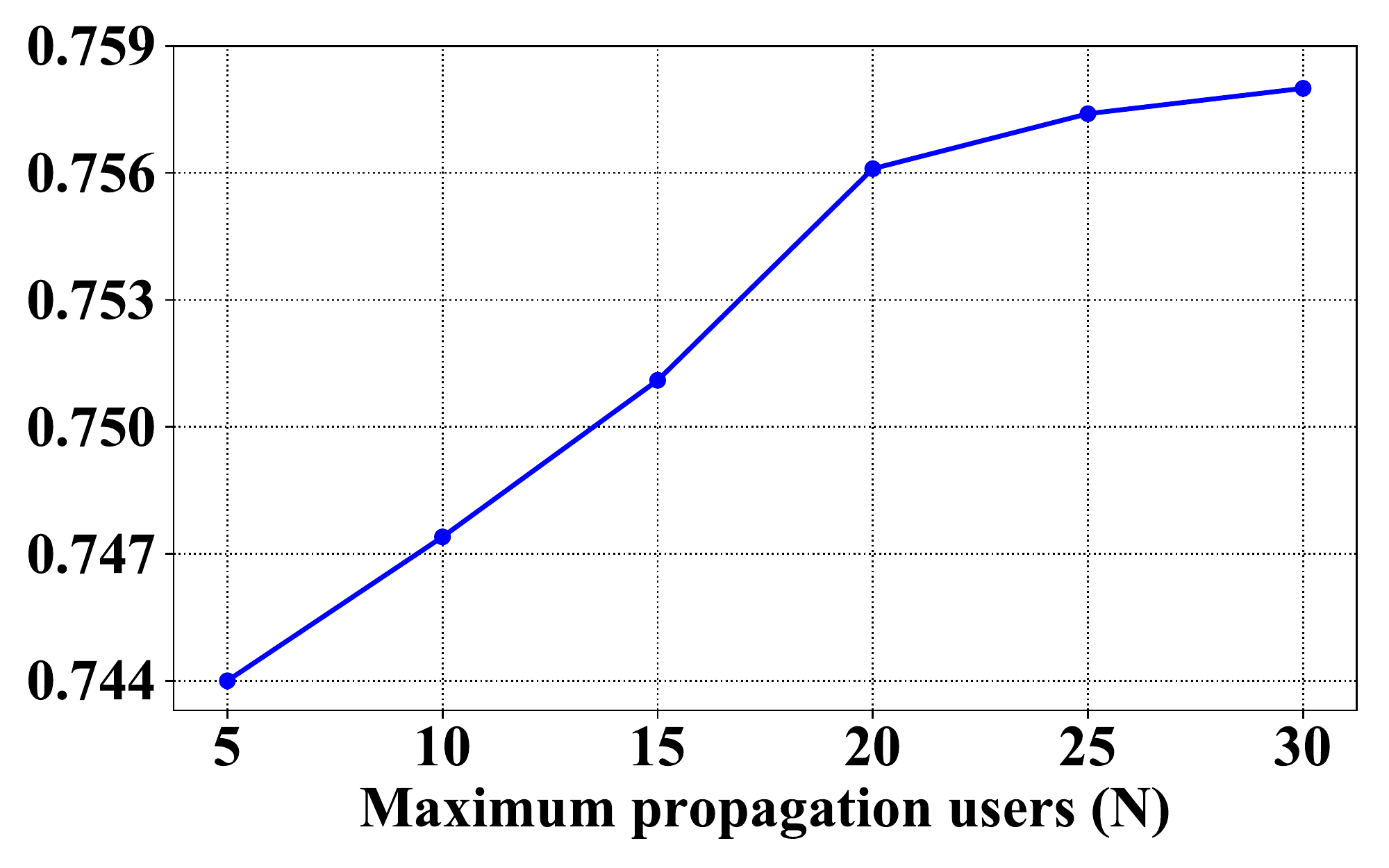}}
\subfigure [\texttt{PriRec} on \emph{Koubei90}] { \includegraphics[width=5cm]{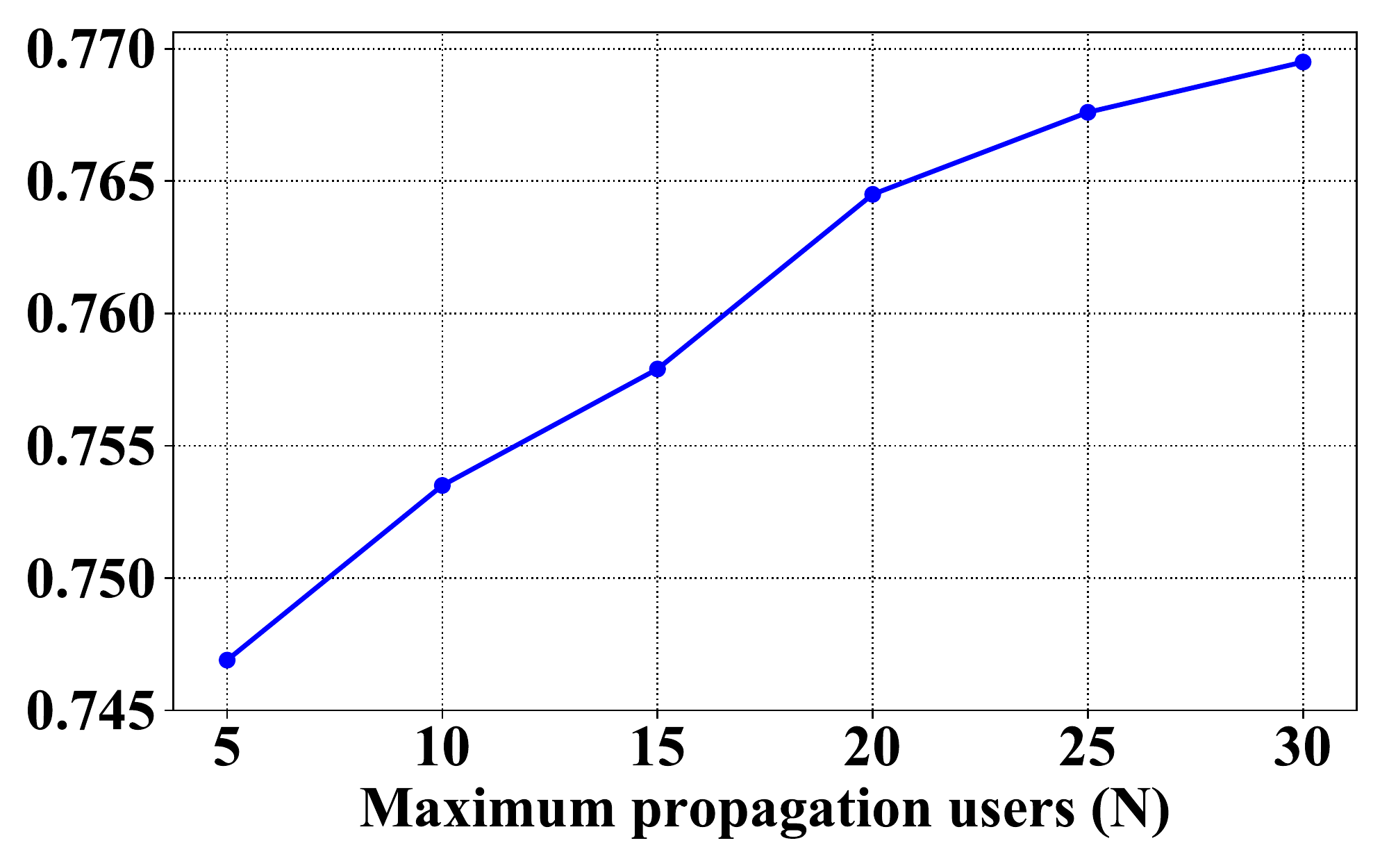}}
%\vskip -0.15in
\caption{Effect of the number of maximum neighbors ($N$) on the AUC of \texttt{PriRec-} and \texttt{PriRec}.}
\label{effectn}
\end{figure*}

Next, we study the effect of the number of maximum neighbors ($N$) on \texttt{PriRec-} and \texttt{PriRec}, which is shown in Figure \ref{effectn}, where we set $K=5$. 
From it, we find that with the increase of $N$, the performances of \texttt{PriRec-} and \texttt{PriRec} first increases and then tends to be stable. 
It indicates that \texttt{PriRec-} and \texttt{PriRec}, without and with POI dynamic features respectively, can achieve stable performance with only a handful of neighbors ($30/85,466 \approx 0.04\%$) to communicate, which meets the situations that only a small proportion of devices are online in practice. 
This experiment proofs the practicalness of our proposed models. 

\begin{figure}[t]
\centering
\includegraphics[width=8cm]{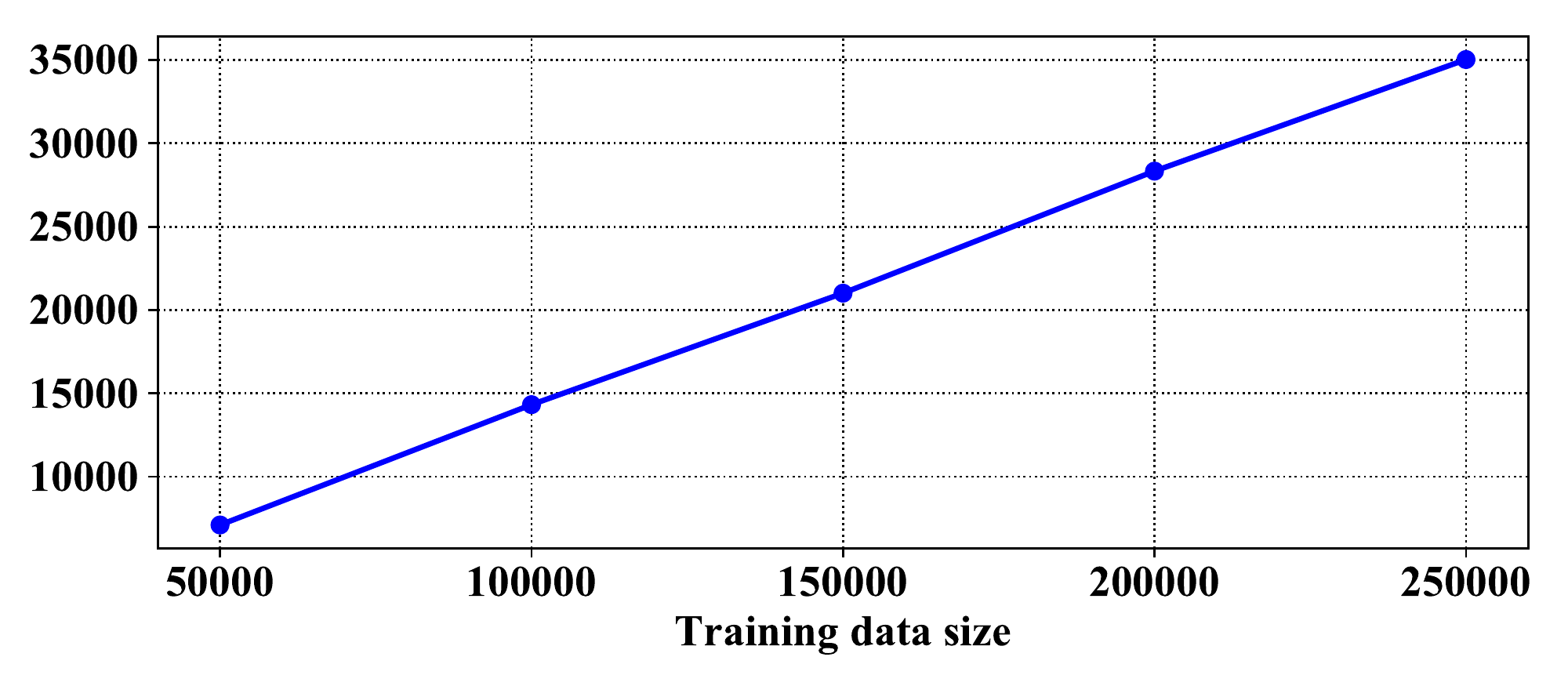}
%\vskip -0.15in
\caption{Training time (in seconds) of \texttt{PriRec} w.r.t. training data size.}
\label{time}
%\vskip -0.15in
\end{figure}

Finally, we study the complexity of \texttt{PriRec}. 
We show the training time of \texttt{PriRec} w.r.t. the training data size in Figure \ref{time}, where $T=100$ and $N=30$. 
Note that our experiments are conducted on a single PC, thus the network communication time is ignored. 
From it, we find that the time complexity of \texttt{PriRec} is indeed linear with training data size, as we analyzed in Section \ref{sec-complexity}, which proofs the efficiency of \texttt{PriRec}. 

\section{Conclusion and Future Work}
In this paper, we proposed a novel privacy preserving POI recommendation (\texttt{PriRec}) framework for the POI recommendation channel in Ant Financial. %, to protect user privacy, including user private profiles and action histories. 
To do this, \texttt{PriRec} keeps users' private profiles on their own devices, and adopts local differential privacy technique to collect perturbed user-POI interaction data on server for generating dynamic POI popularility features. 
Motivated by Factorization Machine (FM), our proposed model of \texttt{PriRec} includes two parts: 
(1) the linear models that are decentralized on each users' side for privacy purpose, which are learnt collaboratively by our proposed secure decentralized gradient descent protocol, and 
(2) the feature interaction model that is kept by the recommender, which is learnt by secure aggregation strategy in federated learning paradigm. 
\texttt{PriRec} not only can protect data and model privacy, but also enjoys promising scalability.  
We applied \texttt{PriRec} in real-world datasets, and comprehensive experiments demonstrated that, compared with FM, PriRec achieves comparable or even better recommendation performance. 

In the future, we would like to deploy \texttt{PriRec} in real products. 
We will also study how to consolidate our algorithm protect against malicious adversary. 
%compress models to save communication cost without lossing much accuracy. 

\bibliographystyle{ACM-Reference-Format}
\bibliography{prirec}

\end{document}